\documentclass[11pt, a4paper]{article}

\usepackage{authblk}

\usepackage[a4paper, total={6.2in, 8.4in}]{geometry}

\usepackage{makeidx}         
\usepackage{graphicx}        
\usepackage[bottom]{footmisc}

\begin{document}

\title{Simple indicators for Lorentzian causets}

\author[1]{Tommaso Bolognesi}
\author[2]{Alexander Lamb}
\affil[1]{CNR/ISTI, Pisa - t.bolognesi@isti.cnr.it}
\affil[2]{Santa Cruz, CA - alex.lamb@gmail.com}
\date{}                     
\setcounter{Maxaffil}{0}
\renewcommand\Affilfont{\itshape\small}


\maketitle

\begin{abstract}
Several classes of directed acyclic graphs have been investigated in the last two decades, 
in the context of the Causal Set Program, in search for good discrete models of spacetime.  
We introduce some statistical indicators that can be used for comparing these graphs 
and for assessing their closeness to the ideal Lorentzian causal sets ('causets') -- those obtained by \emph{sprinkling} points in a Lorentzian manifold.  
In particular, with the reversed triangular inequality of Special Relativity in mind, 
we introduce 'longest/shortest path plots', an easily implemented tool to visually detect 
the extent to which a generic causet matches the wide range of path lengths between events of Lorentzian causets.
This tool can attribute some degree of 'Lorentzianity' - in particular 'non-locality' - also to causets that
are not (directly) embeddable and that,  due to some regularity in their structure,
would not pass the key test for Lorentz invariance: the absence of preferred reference frames.
We compare the discussed indicators and use them for assessing causets 
both of stochastic and of deterministic, algorithmic origin, finding examples of the latter 
that behave optimally w.r.t. our longest/shortest path plots.
\end{abstract}

%
%
\section{Introduction}
\label{sect:Introduction}

A rather direct way to obtain a discrete model of spacetime from a continuous one 
- i.e. from a Lorentzian manifold that satisfies the Einstein field equations - 
consists in applying to the latter the \emph{sprinkling technique} and building a \emph{causal set}.


A \emph{causal set}, or \emph{causet} \cite{ref:bombelli87, ref:SorkinValdivia2003, ref:rideoutsorkin}, 
is a partially ordered set ('poset') of atomic spacetime events in which the partial order relation '$\preceq$'
expresses causal dependencies between events.  
Beside being reflexive, antisymmetric and transitive, relation '$\preceq$' 
is required to be \emph{locally finite}, or \emph{finitary}, that is, for any pair $(s, t)$ of events, the set of points between them must be finite: 
$|\{x | s\preceq x \preceq t\}| < \infty$.  

A causet can be represented by a DAG (Directed Acyclic Graph), whose edges reflect the partial order relation.  
Any such DAG is transitively closed, by definition, but we shall also consider \emph{transitive reductions} of these graphs (their 'Hasse' diagram), from which the original causet is recovered by transitive closure.  
The edges of a causet - of the transitively closed DAG that represents it - are often called 'relations', 
while those of its transitive reduction are  called 'links', and identify the nearest neighbours of each event/node.  

In light of our interest here for discrete spacetime modeling, in the sequel we shall sometimes abuse the term 'causet'
and sloppily use it even when the generic term 'DAG' would be more appropriate.  
For example, we shall consider procedures that build "causets" which are neither closed nor reduced - these are called 'raw' causets.  
The statistical indicators that we discuss in the paper, however, 
shall always refer either to the transitive closure or to the transitive reduction of these graphs.

By the \emph{sprinkling technique}  \cite{ref:bombelli87, ref:SorkinValdivia2003, ref:rideoutsorkin} one can 
derive a causet from a  Lorentzian manifold $M$ provided with a volume measure, in two steps.
First one considers a uniform, Poisson distribution of points - to become the causet nodes - in a finite region of $M$, with density $\delta$,
so that the expected number of points in a portion of volume $V$ is $\delta V$
and the probability to find exactly $n$ points in that portion is:
\begin{equation}
P(n) = \frac{(\delta V)^n e^{-\delta V}}{n!},
\end{equation}
Then the causet edges are created by letting the sprinkled points inherit the causal (lightcone) structure of $M$.
In the sequel we shall conveniently call these objects \emph{sprinkled causets}. 

A challenging goal of causet-based quantum gravity research is to reverse the above logic, and to try and build causets of physical significance without resorting to an underlying continuum.  Several techniques for doing this have been investigated in the last two decades, and some of them will be described in Section \ref{sect:CausetConstructionTechniques}.  Under this perspective, the manifold is only obtained a posteriori, as an asymptotic approximation.   

One way to assess a causet $C$ obtained by some of these techniques is to check whether 
$C$ is \emph{faithfully embeddable} in some manifold $M$ of appropriate dimension, that is, whether it could have arisen with high likelihood from sprinkling events in $M$.  
In \cite{ref:Henson2006} Henson introduces a method for building an actual embedding of a given causet in 2D Minkowski space, but recognizes the need of additional,  complementary, possibly more efficient criteria for defining scales of causet 'manifoldlikeness'.  Some of these efforts are mentioned in the conclusive section.

In the approach described here, we are interested in extracting some distinguishing feature of sprinkled causets 
\emph{other than}, or 
\emph{weaker than faithful embeddability}. 
One simple reason is that embeddability is hard to test; furthermore, it might fail on the original causet while succeeding with a coarse-grained version of it.  
Hence we shall try to identify statistical features, in the discrete setting, that could reflect the peculiarities of the Lorentz pseudo-metrics in the continuum, which is responsible for the reversed triangular inequality and the twin paradox of Special Relativity.  
We shall then use these indicators as a benchmark for the graphs obtained by other techniques.  
While it is today sufficiently understood that '\emph{a fundamental spacetime discreteness need not contradict Lorentz invariance}' \cite{ref:DawkerHensonSorkin2004}, 
the toolkit for assessing causets in terms of Lorentzianity is still quite poor. 

In Section \ref{sect:CausetConstructionTechniques} we briefly recall four stochastic causet construction techniques, including sprinkling in flat (Minkowski) and positively curved (de Sitter) manifolds.

In Section \ref{sect:CountingEdges} we concentrate on statistical properties based on counting \emph{relations}.  
Keeping in mind that sprinkled causets are \emph{transitively closed} by construction, the counts shall be concerned with the edges of 
the transitively closed versions of the graphs obtained by the other techniques.  
First we analyse the distributions of \emph{node degrees} - the number of emanating relations -, 
finding that power-law distributions are quite common.
Then we collect other statistical information relevant to causet dimensionality estimation, 
and visualise it in what we call \emph{ordering fraction spectra}.

In Section \ref{sect:CountingLinks} we focus on the \emph{transitive reduction} of the considered graphs, thus we count \emph{links}.
This turns out to be a more discriminative criterion for addressing Lorentzianity issues.  
We first analyze the (link-)degree of the root node of an interval
as a function of the number of nodes in the interval.  Then we introduce a special type of diagram - longest/shortest path plots - meant to expose the peculiar wide range of path lengths between two generic nodes, a feature of Lorentzian sprinkled graphs sometimes referred to as 'non-locality'.  We illustrate, again, how the various causet classes perform with respect to these indicators.

In Section \ref{sect:PermutationAnts} we introduce \emph{deterministic} causet construction techniques
based on permutations of tuples of natural numbers and on their manipulation by stateless or stateful control units; 
we show that one instance of this model (out of 65,536) yields a pseudo-random causet with 
excellent non-locality properties comparable to those of sprinkled causets.

In Section \ref{sect:Conclusions} we summarize our results and mention some related work.

Let us stress again that the proposed indicators reflect a concept of 'Lorentzianity'  
weaker than faithful embeddability in a Lorentzian manifold; 
they \emph{do not provide necessary and sufficient conditions} for this strongest form of Lorentzianity.   
The idea is to investigate other properties of sprinkled Lorentzian causets (embeddable by definition), 
and select some that can be regarded as sharply characterising  these causets of reference.

%
%
\section{Stochastic causet construction techniques}
\label{sect:CausetConstructionTechniques}

In this section we introduce the four stochastic causet construction techniques to be analyzed in the rest of the paper.  In addition, and for the sake of comparison, we shall occasionally consider regular lattices.  
In the sequel we use the concept of \emph{order interval}, often abbreviated to 'interval': 
if $s$ and $t$ are two points of a discrete or continuous set with partial order $\preceq$, 
then the order interval between them, denoted $I[s, t]$, is the set 
$\{x | s\preceq x \preceq t\}$, which includes $s$ and $t$, by the reflexivity of the order relation.

%
\subsection{Minkowski sprinkling}
\label{subsect:MinkowskiSprinkling}

A Minkowski sprinkled causet is obtained by applying the already mentioned sprinkling technique \cite{ref:bombelli87, ref:SorkinValdivia2003, ref:rideoutsorkin} 
to a portion of $d$-dimensional Minkowski space.  
For example, if $M^{(1,2)}$ is 3D Minkowski space with time dimension $t$ and space dimensions $x$ and $y$, 
$L^{2}$ is the squared Lorentz distance 
$L^{2}(p(t_{p}, x_{p}, y_{p}), q(t_{q}, x_{q}, y_{q}) =  +(t_{p} - t_{q})^2 -(x_{p} - x_{q})^2 -(y_{p} - y_{q})^2$ , 
and $C$ is a bounded, connected subset of it, e.g. a unit cube, 
then a set $S$ of points Poisson-distributed in $C$ yields a sprinkled causet graph $G(S, E)$, 
where the set of graph nodes is $S$ itself and the edges $E$ are the ordered pairs of nodes $(p, q)$ 
such that $q$ is in $p$'s future lightcone: 
$E = \{(p, q) \in S^{2}) | L^{2}(p, q) \geq 0 \wedge t_{p} < t_{q}\}. $
Once the graph is constructed, node coordinates become irrelevant.  
In general $G(S, E)$ may have several sources - nodes whose in-degree is zero - and sinks - nodes whose out-degree is zero; 
however, given two nodes $s$ and $t$ of it, 
the order interval $I[s, t]$ has, by definition, only one source ($s$) and one sink ($t$).

%
\subsection{De Sitter sprinkling}
\label{subsect:DeSitterSprinkling}

De Sitter spacetime is an exact solution of the Einstein field equations of General Relativity, 
conjectured to model the universe both at the Plank era (at time $t < 10^{-44}$ sec.) and in the far future, at thermodynamic equilibrium.  
De Sitter spacetime has constant positive curvature, and describes the universe as an empty and flat space 
- one with zero curvature - which grows exponentially under the effect of a positive cosmological constant.  

Sprinkling in de Sitter spacetime, up to four dimensions, is discussed in \cite{ref:Krioukov2012},
where various properties of the corresponding causets are studied both analytically and by simulation,
and compared with properties of various other types of network.

Here we restrict to two-dimensional de Sitter spacetime $DS^{(1,1)}$, for which a 
well known representation as a one-sheeted, 2D hyperboloid embedded in flat 3D Minkowski space $M^{(1,2)}$ is available,
and, following in part \cite{ref:Krioukov2012}, we present the simple analytical steps behind the implementation
of 2D de Sitter sprinkling we have used for our experiments.

Let $t$, $x$, $y$ be the axes of $M^{(1,2)}$, where $t$ is the vertical time axis around which a hyperbola rotates to create the hyperboloid, 
and let $\tau$ and $\theta$ be the time and space coordinates in $DS^{(1,1)}$: in the embedding, $\tau$ flows vertically 
while angular coordinate $\theta$ spans space - a circle of constant coordinate $t$ (and $\tau$).
Given the hyperboloid equation $x^{2}+ y^{2}- t^{2} = 1$, 
the differential element of Lorentzian length  $ds^{2}= dt^{2}- dx^{2}- dy^{2}$, 
and its vertical projection $d\tau^{2}= dt^{2}- dx^{2}$, 
one can readily derive, by integration of the latter along a vertical hyperbolic segment, 
the functional dependences $\tau(t) = ArcSinh(t)$ and $t(\tau) = Sinh(\tau)$.  
Similarly, the radius of the space circle at time $\tau$ is $r(\tau) = Cosh(\tau)$, hence the size of de Sitter space is 
$2\pi Cosh(\tau) = \pi(e^{\tau}+e^{-\tau}).$
Thus, space grows exponentially with time.  
(For a visualization of this result, see the interactive demo \cite{ref:BolognesiDeSitterDemo}.)  
It is finally easy to see that the differential element of Lorentzian length in $DS^{(1,1)}$ is $ds^{2}= d\tau^{2}- Cosh(\tau)d\theta^{2}.$ 

For implementing de Sitter sprinkling we need to know how to distribute points on the surface of the hyperboloid, 
and how to build causal edges.  
If $\delta$ is the desired uniform density, the expected number of points on the circular hyperboloid section $S$ 
between $\tau$ and $\tau + d\tau$ is $\delta Area(S) =  \delta 2\pi Cosh(\tau)d\tau$.  
Thus, if we uniformly sprinkle in a section of the hyperboloid with $0\leq \tau \leq \tau_{max}$,
the time coordinate of these points is a random variable $\mathbf{\tau}$ whose (normalized) density is: 
\begin{equation}
f_{\mathbf{\tau}}(x) = \mbox{\emph{Cosh}}(x)/ \mbox{\emph{Sinh}}(\tau_{max}),
\end{equation}
(as already established in \cite{ref:Krioukov2012}, equation (3)), where we use the fact that $\int_{0}^{\tau_{max}} Cosh(x) dx = Sinh(\tau_{max})$.

The Minkowski coordinate $t$ of these points is itself a random variable $\mathbf{t}$.  
Keeping in mind the above functions $t(\tau)$ and $\tau(t)$, and using a fundamental theorem on functions of random variables, 
we find that the density $f_{\mathbf{t}}$ of $\mathbf{t}$ is constant:
\begin{equation}
f_{\mathbf{t}}(x) = \frac{f_{\mathbf{\tau}}(\tau(x)))}{t'(\tau(x))} = \mbox{\emph{Csch}}(\tau_{max}),
\end{equation}
where '\emph{Csch}' is the hyperbolic cosecant.
Then, in light of the circular symmetry of the distribution, implementing a Poisson sprinkling in $DS^{(1,1)}$, 
with $\tau$ ranging in $[0, \tau_{max}]$, is straightforward. 
We create points with coordinates $(r, \theta, t)$, where polar coordinates $(r, \theta)$ replace $(x, y)$, such that:

\begin{itemize}
\item $t$ is distributed uniformly in $[0, t_{max}]$, where $t_{max} = \mbox{\emph{Sinh}}(\tau_{max})$,
\item $r = \sqrt{t^{2} + 1}$,
\item $\theta$ is distributed uniformly in $[0, 2\pi)$.
\end{itemize}

Once the points are uniformly distributed in $DS^{(1,1)}$ as described, 
causal edges can be established among them by referring directly to their Lorentz distance in $M^{(1,2)}$: 
although the squared Lorentz distance changes when moving from $p$ to $q$ on a geodesic in$DS^{(1,1)}$
or on one in $M^{(1,2)}$, the signs of these two distances always agree, thus yielding the same causal structure.

%
\subsection{Percolation dynamics}
\label{subsect:PercolationDynamics}
%
Transitive percolation dynamics has been widely studied in the context of the Causal Set Programme 
\cite{ref:rideoutsorkin, ref:AhmedRideout2010, ref:rideout2004, ref:RideoutSorkin2001}.
An $N$-node percolation causet is built by progressively numbering nodes 1, 2, ..., $N$, 
and by creating an edge $i \rightarrow j$, for $i <  j$, with fixed probability $p$ (typically $p \ll 1$).  
Note that a percolation causet may be disconnected and may include several source nodes.  
A variant called \emph{originary percolation}, which guarantees the existence of a 
single source node, or \emph{root}, is studied in \cite{ref:AhmedRideout2010}, 
where it is also observed that any subset $S$ of a standard, transitive percolation causet $C$, 
consisting of a node $x$ and all nodes in its future, is an instance of an originary percolation causet. 

%
\subsection{Popularity/Similarity dynamics}
\label{subsect:PopularitySimilarityDynamics}
%
This technique is described in \cite{ref:Krioukov2012}, where it is also shown that the growth dynamics is asymptotically identical 
to that of sprinkled causets from de Sitter space.  
Each node is assigned a progressive natural number $n$, as in percolation dynamics, 
and is placed in 2D Euclidean space, with polar coordinates $(r, \theta)$, 
where $r$ is a monotonic, increasing function of $n$ (whose precise nature is not essential for building the causet)
and $\theta$ is a random angle in $[0, 2\pi)$.  
When new node $n$ is created, a fixed number $m$ of edges reaching $n$ from previously created nodes is added: 
$s_{i} \rightarrow n$, with $s_{i} \in \{1, 2, \dots (n-1)\}$ and $i$ = 1, 2, ..., $m$.
Typically, $m$ = 2, and this is the value we use throughout the paper, unless otherwise stated.
Nodes $s_{i}$ are those that minimize the product 
$s_{i} \Delta \theta_{i}$,
where $\Delta \theta_{i}$ is the angular distance of node $s_{i}$ from node $n$.  

The name 'popularity/similarity', abbreviated as 'pop/sim' in the sequel, reflects the tradeoff between popularity and similarity 
that determines the structure and drives the growth of various complex networks, including the Internet
\cite{ref:Papadopoulos2012}.
Node labels (birth dates) represent 'popularity', while angular distances express 'similarity': 
when the from-nodes $s_{i}$ are chosen for being connected to new node $n$, 
older nodes - with smaller labels - are preferred, and in the long run they become more and more 'popular', 
thus increasing their out-degrees; 
at the same time, the chosen nodes must be as 'similar' as possible (small angular distance) to the target node $n$.  

%
\subsection{Regular grids}
\label{subsect:RegularGrids}
%
For the sake of comparison we shall also consider directed regular grids, in particular square grids 
whose edges are oriented parallel to the cartesian axes of their embedding euclidean 2D space.  
These graphs manifest maximum locality, as opposed to the non-locality typical of sprinkled causets from a Lorentzian manifold.  

Randomized versions of maximally local graphs can be obtained by sprinkling points in Euclidean space of some dimension, 
where one of the dimensions is time, and by creating an edge from node $p$ to node $q$ whenever $q$ has 
higher time coordinate than $p$ and the Euclidean distance between $p$ and $q$ is smaller than a given threshold.

%
\section{Counting relations}
\label{sect:CountingEdges}

In this section we try to characterize causets by collecting statistical information based on edge counts for transitively closed graphs.  
We focus on two types of statistical indicator: \emph{node degree distributions} and \emph{ordering fraction spectra}.  
These are applied to our reference causets - sprinkled Minkwski and de Sitter - and to the other causet classes: 
since the former are transitively closed by construction, 
we shall take transitive closures of the latter too, before counting edges.

%
\subsection{Node degree distributions}
\label{subsect:NodeDegDistribForTransClosedCausets}
%
Minkowski space $M^{(1,1)}$ represents a spacetime that extends to infinity in both dimensions; 
a finite causet is obtained by sprinkling points in a bounded region of it.  
However, depending on the observed variables, 
the statistics of such causets may be affected by the shape of the region border.  
One way to avoid this problem is to analyze causet \emph{intervals}, as defined above,
although we shall not ignore other types of causets whose analysis we find useful,
for reasons to be explained as these are introduced.

%
\subsubsection*{Intervals: sprinkled Minkowski, square grid, sprinkled de Sitter}
Thus, let us start by investigating the distributions of node degrees for \emph{causet intervals}, 
where the degree of a node is the number of edges emanating from it (out-degree).  
We shall represent these distributions by histograms, 
built by segmenting the range of possible degrees into bins of fixed or variable width.

\begin{figure}[h]
\centering
\includegraphics[scale=.6]{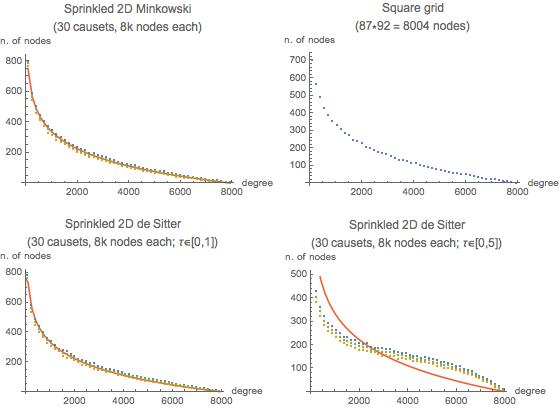}
\caption{{\footnotesize 
Node degree histograms for transitively closed interval causets. 
Each histogram refers to 50 slots of equal width. 
For the three sprinkled cases the histogram was obtained by averaging over thirty interval causets;
for each slot, standard deviations from the mean node count are also shown.
To reduce graphical cluttering the vertical bar representation of the bins is avoided.
Solid lines represent fitting negative logarithm functions. }}
\label{fig:TrCloNodeDegsIntervalCausets}     
\end{figure}

The four plots in Figure \ref{fig:TrCloNodeDegsIntervalCausets} 
show histograms for the node degrees of four types of interval causet.
%
%
%
%
The histograms contain 50 points each; 
they were obtained by partitioning the range of possible degrees into 50 slots of equal width 
and by counting the nodes whose degrees fall in each slot.
For the three cases corresponding to sprinkling in Lorentzian manifolds (Minkowski and de Sitter),
each histogram has been obtained by averaging over 30 causets.  
For each slot the plot provides \emph{three} points, that indicate  the averaged
value of the node count and the standard deviation.

In the upper-left plot, referring to thirty 8000-node sprinkled 2D Minkowski intervals, 
the continuous line represents the theoretical density $-Log(z)$, appropriately scaled.  
This negative logarithm density can be derived by assimilating the node degree to a random variable (r.v.) 
$\mathbf{r}=\mathbf{x}  \mathbf{y}$, 
where r.v.'s $\mathbf{x}$ and $\mathbf{y}$ have uniform densities in the unit interval: 
$f_{\mathbf{x}}(z) = 1$, $f_{\mathbf{y}}(z) = 1$, $z \in [0,1]$.  
The reason is that, by a -45 degree rotation, we can represent the sprinkled Minkowski interval 
as a unit square $[0, 1] \times [0, 1]$ in Euclidean 2D space, 
and model the degree of the generic sprinkled point $p(1-x, 1-y)$ as the area $xy$ of the rectangle of dimensions $x$ and $y$.  
The distribution function for $\mathbf{r}$ is: $F_{\mathbf{r}}(z) = Prob[\mathbf{x}\mathbf{y} \leq z] = z(1-Log(z))$, 
yielding, by derivation, the density function $f_{\mathbf{r}}(z) = -Log(z)$.  

The analysis above is essentially valid also for the 8004-node regular square grid - an interval too -
whose node degree histogram is shown in the upper-right plot. 
(The grid has integer node coordinates, source $s(1, 1)$, sink $t(92, 87)$, and edges oriented upward or to the right.)

Each of the two plots in the lower row of Figure \ref{fig:TrCloNodeDegsIntervalCausets} 
refers to thirty 8000-node sprinkled 2D de Sitter intervals, with time parameter $\tau$ ranging, respectively,
in $[0, 1]$ and $[0, 5]$.  
For a small $\tau_{max}$ the de Sitter and Minkowski intervals are quite similar, as reflected in their node degree distributions. 
As the width of the time window increases, the effect of curvature becomes more sensible and the plot departs
from the logarithmic behaviour.

A first conclusion can be already drawn by a qualitative analysis of the plots in 
Figure \ref{fig:TrCloNodeDegsIntervalCausets}:
node degree distributions of transitively closed graph intervals fail to characterise Lorentzianity,
since they do not distinguish between a sprinkled Minkowski or de Sitter interval -- the Lorentzian causets of reference -- 
and a regular grid, which is the typical example of a \emph{non}-Lorenzian causet.


\subsubsection*{Full circular causets: sprinkled Minkowski, square grid, sprinkled de Sitter}

In spite of the poor performance of node degree distributions as Lorentzianity indicators,
relative to causet \emph{intervals}, 
we are interested in investigating a bit further their application to \emph{full} causets. 
This step is mainly suggested by the observation that the infinite hyperboloid representation of 
2D de Sitter spacetime in 3D Minkowski spacetime (subsection \ref{subsect:DeSitterSprinkling}) lends itself
to an additional, natural 'cutting' operation for obtaining finite portions of the manifold, beside the extraction 
of diamond-shaped intervals, namely
the selection of the \emph{full slice} of the curved surface 
between two parallel, horizontal planes in the embedding 3D Minkowski spacetime,
with time parameter $\tau$ ranging between, say, $0$ and $\tau_{max}$. 
Unlike in an interval, this full slice spans (finite) space completely.
(In Minkowski spacetime we need to explicitly bound both time and space, for getting a finite region.)
These precise full slices of de Sitter spacetime have been considered in \cite{ref:Krioukov2012},
where degree distributions are found to follow a power law.  
We are interested
in comparing those results with \emph{cylindrical counterparts} of sprinkled Minkowski and square grid causets,
for further assessment of the indicator.

\begin{figure}[h]
\centering
\includegraphics[scale=.55]{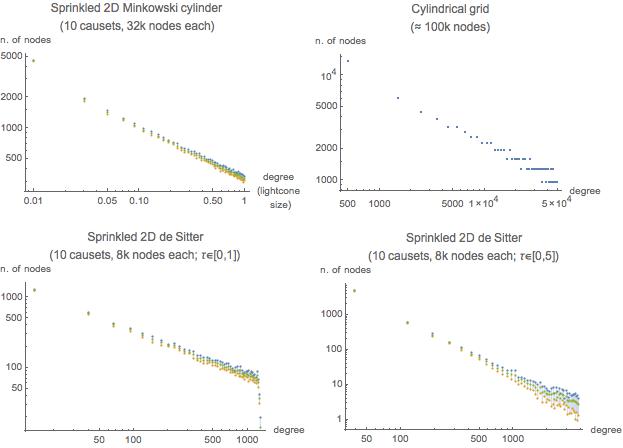}
\caption{{\footnotesize Node degree histograms for transitively closed causets with rotational symmetry.
Similar to Figure  \ref{fig:TrCloNodeDegsIntervalCausets}, each histogram refers to 50 slots of equal width.
For the three sprinkled cases the histogram was obtained by averaging over ten interval causets; 
standard deviations from the mean node count are also shown.
\emph{LogLog} plots are used here, for highlighting the power law character of the distributions.}}
\label{fig:TrCloNodeDegsRotationalCausets}     
\end{figure}

Thus, we have considered two full slices of de Sitter spacetime,
with time bounds $[0,1]$ and $[0, 5]$, as those used in Figure \ref{fig:TrCloNodeDegsIntervalCausets},
and for each of them we have derived ten sprinkled, 8k-node causets, and the corresponding node degree histograms,
where the range of possible degrees has been partitioned into 50 slots of equal size.
These are shown in the lower row of Figure \ref{fig:TrCloNodeDegsRotationalCausets}.
The two \emph{LogLog} plots confirm the power-law character of these distributions. 

%


Inspired by the rotational symmetry of the two de Sitter full slices,
\footnote{All four causet types presented in Figure \ref{fig:TrCloNodeDegsRotationalCausets} 
can be though of as embedded in a 3D space:
the term 'rotational symmetry' refers to this external viewpoint.  Under an internal viewpoint, the symmetry is translational. 
}
and for the sake of comparison, 
in the upper row of Figure \ref{fig:TrCloNodeDegsRotationalCausets}
we have considered the cylindrical counterparts
of the interval causets covered in the upper row of Figure \ref{fig:TrCloNodeDegsIntervalCausets} - sprinkled Minkowski and square grid.

The upper-left plot of Figure \ref{fig:TrCloNodeDegsRotationalCausets}
shows the node degree histogram for ten causets obtained by uniformly sprinkling points in a rectangular portion of $M^{(1,1)}$ 
that wraps completely around a cylinder of unitary height - a 'Minkowski cylinder'.  
\footnote{
This is essentially analogous to sprinkling in a finite, rectangular, spacelike strip of $M^{(1,1)}$,
with the advantage of avoiding the (slight) boundary effects on the statistics induced by the two vertical cuts. }
The out-degree of a node in the Minkowski cylinder causet is equated to the area of the future lightcone of that node, 
whose value ranges from 0 to 1. 
The distance $\mathbf{h}$ of the generic node $p$ from the upper border of the cylinder is a r.v. with uniform density: 
$f_{\mathbf{h}}(x) = 1$, for $x \in [0,1]$.  
Assuming that the cylinder radius is a large enough to prevent any future lightcone to overlap with itself, 
the area of $p$'s future lighcone is r.v. $\mathbf{A} = h^{2}$.  
By applying well known results on functions of r.v.'s, we obtain $f_{\mathbf{A}}(y) = \frac{1}{2} y^{-1/2}$, $y \in [0, 1]$ 
for the density of r.v. $\mathbf{A}$.  
Thus the density is a power-law, and the \emph{LogLog} plot in Figure  \ref{fig:TrCloNodeDegsRotationalCausets} (upper-left) 
is correspondingly linear.  
In the histogram, node degrees are partitioned into 50 slots of equal width. 

The upper-right plot of Figure \ref{fig:TrCloNodeDegsRotationalCausets}
refers to a square grid arranged around a cylinder ('cylindrical grid'), with edges at +45 and -45 from the cylinder axis - 
a regular counterpart of the sprinkled Minkowski cylinder.  
The analysis of the node degree density is somewhat analogous to that of the Minkowski cylinder.  
Similar to the previous case, the histogram is obtained by partitioning the degrees into 50 equal slots, 
and reveals a power-law distribution.  
The vertical quantisation is due to the regularity of the graph, which is formed by $2n-1$ rows of $2n-1$ points each, yielding a total of 100,489 nodes, for our choice of $n = 159$.  Note that nodes are aligned vertically only at alternate rows.  
All 317 nodes of each row have the same degree, 
hence the populations of degrees in each of the 50 slots of the histogram necessarily differ by multiples of 317.

\subsubsection*{Full pop/sim causet} 

\begin{figure}[h]
\centering
\includegraphics[scale=.6]{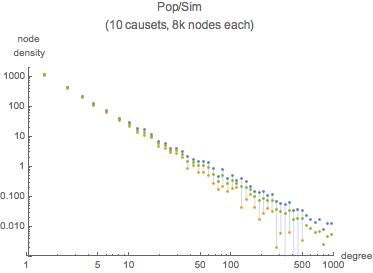}
\caption{{\footnotesize Node degree histogram for ten, 8k-node transitively closed pop/sim causets. 
Bins of geometrically increasing width have been used, and node degree values have been normalised
accordingly, providing a density that may be smaller than 1.
Standard deviations from the mean are shown. Lower standard deviations for some of the rightmost points
fall below zero, thus escaping representation  on the \emph{Log} scale.
}}
\label{fig:TrCloNodeDegsPopSim}     
\end{figure}

The histogram in Figure \ref{fig:TrCloNodeDegsPopSim} refers to
ten (transitively closed) 8k-node pop/sim causets, with fixed node in-degree $m = 2$.
For improved smoothness we have used geometrically increasing bins, thus the number of nodes for each slot has been divided by the bin width,
providing a degree \emph{density}.  Again, a \emph{LogLog} plot was chosen for highlighting that a power law is in action. 

%

Note that the pop/sim causet, described in terms of polar coordinates $(r, \theta)$, 
could also be regarded as rotationally symmetric; 
one difference with the previous causets of this type, however, is that the 
$r$ 'coordinate' plays here a somewhat abstract role, being used purely for ordering nodes as they are created, 
not for defining a particular metric of the manifold where sprinkling takes place.

The fact that the plot in Figure \ref{fig:TrCloNodeDegsPopSim} reflects a power law, 
like the plots in Figure \ref{fig:TrCloNodeDegsRotationalCausets},
is not surprising since
the growth dynamics of the pop/sim procedure, relative to node degree distribution, is shown in \cite{ref:Krioukov2012} to be
asymptotically identical to that of de Sitter sprinkled causets.


\vspace{0.5cm}

The results just illustrated for full, circular causets neatly confirm the conclusion we have anticipated
about node degree distributions as potential Lorentzianity indicators:
these distributions do not discriminate between sprinkled causets - our reference - and grids,
neither when looking at intervals (Figure \ref{fig:TrCloNodeDegsIntervalCausets})
nor when considering full circular instances 
(Figures \ref{fig:TrCloNodeDegsRotationalCausets} and \ref{fig:TrCloNodeDegsPopSim}). 

Having ruled out this indicator, we would not need to further analyse the only case left - percolation causets.
Yet, the analysis of their degree distributions may be of some interest in itself, 
due to the peculiar role played by the edge probability parameter,
and is presented in Appendix A.
Interestingly, a power law distribution essentially emerges also in that case, providing further
evidence of the 'flattening' effect of transitive closure, an operation which seems to obscure 
important structural differences among 'raw' causets of different kinds.
\footnote{
Indeed, the analysis and comparisons in \cite{ref:Krioukov2012} deal also with the actual power law exponent $-\gamma$.
For the de Sitter universe, $\gamma$ is found to evolve on a cosmological time scale from value 3/4 to an asymptotic value 2.  
The latter value is found to characterise also several networks of different origin.
}

%
\subsection{Ordering fraction spectra}
\label{subsect:OrderingFractionSpectra}
%
%
Counting the relations of causet intervals
is also at the basis of the Myrheim-Meyer dimension estimator.  
Let $I_{k}^{D}[s,t]$ 
be a $k$-node interval with source $s$ and sink $t$, obtained by sprinkling in $D$-dimensional Minkowski space.  
As $k$ grows, the expected number $R(D, k)$ of edges in $I_{k}^{D}[s,t]$ quickly approaches (from above) the value \cite{ref:Myrheim1978, ref:SorkinValdivia2003}:  
\[
R(D,k) = f(D){k \choose 2}
\]
where
\[
f(D) = \frac{3}{2 {3D/2 \choose D}}.
\]
In general, the \emph{ordering fraction} of a $k$-node causet (interval) is defined as the ratio 
$R/{k \choose 2}$
between the number $R$ of edges in it and the maximum number 
${k \choose 2}$
of directed edges that could connect so many nodes.  
Thus, the expected ordering fraction of a sprinkled , $k$-node, $D$-dimensional Minkowski interval
must quickly approach $f(D)$, as $k$ grows.
We can then obtain an estimation of the Myrheim-Meyer dimension $D$ of a generic causet 
by counting its nodes ($k$) and edges ($R$) and numerically inverting function $f(D)$:
\[
D = f^{-1}\left(\frac{R}{{k \choose 2}}\right)
\]

\begin{figure}[h]
\centering
\includegraphics[scale=.63]{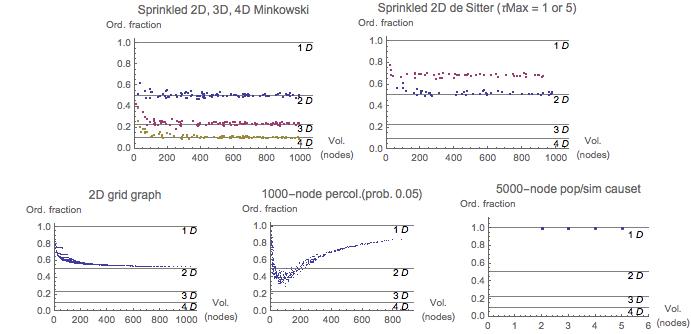}
\caption{{\footnotesize Ordering fraction spectra. Upper row: sprinkled causets.
Lower row: grid, percolation and pop/sim causets.}}
\label{fig:ordFractSpectraForCausetClasses}     
\end{figure}

The upper-left plot of Figure \ref{fig:ordFractSpectraForCausetClasses} 
was obtained by creating 100 intervals of random volume (number of points) lower than 1000 
in 2D-,  3D- and 4D-Minkowski space, and by plotting, for each interval $I$, the point (\emph{Volume}($I$), \emph{OrderingFraction}($I$)).  

As $k$ grows, the points nicely align to the expected values - the horizontal gridlines - which mark the values of $f(D)$ 
for the indicated dimensional values.  
We call these diagrams \emph{ordering fraction spectra}.  
Let us stress that this indicator focuses on \emph{intervals} extracted from graphs of arbitrary kind,
and is therefore unaffected by the overall shape or 'border' (if any) of the latter. 

The upper-right plot of Figure \ref{fig:ordFractSpectraForCausetClasses} 
was obtained by creating 50 intervals of random volume
 -- again less than 1000 points -- in 2D de Sitter spacetime; 
 for the lower set of points in the plot, that matches well the 2D ordering fraction value 0.5, 
 time variable $\tau$ ranges in [0, 1], while for the upper points the range is [0, 5].  
 Note that both point sets are derived from sprinkling in (the hyperboloid model of) 2-dimensional de Sitter spacetime.  
 This plot thus indicates that, while the Myrheim-Meyer dimension estimator operates correctly for flat, Minkowski spacetime, it becomes unreliable when sprinkling in curved manifolds, becoming sensitive to the interval time span. 
 This circumstance was already observed in \cite{ref:MeyerThesis1989}. 
 In any case, based on the above definition of ordering fraction, the flat nature of the plots for sprinkled Minkowski and de Sitter causets implies that in these cases the growth rate of the edge count function $R(k)$ is $O(k^2)$. 
 
In the lower part of Figure \ref{fig:ordFractSpectraForCausetClasses}  
we show three more ordering fraction spectra.   
The first diagram plots the (\emph{Volume}($I$), \emph{OrderingFraction}($I$)) pairs for all intervals $I$ 
of a $32\times32$ directed square grid. 
The ordering fraction value for an $n\times m$ directed square grid interval is:  
 \[
\frac{\sum_{x=1}^{m} \sum_{y=1}^{n} xy -mn}{{mn \choose 2}}.
 \]
 
Similar to the sprinkling case, this plot reveals that the Myrheim-Meyer dimension estimator is valid asymptotically.

The second plot shows the typical shape of the ordering fraction spectra of causets obtained from percolation dynamics 
using constant edge probability.  
The plot was obtained by randomly sampling 500 intervals from the transitive closure of a 1000-node causet with edge probability 0.05.   
On the large scale, percolation causets tend to be one-dimensional.

The poor ordering fraction spectrum in the third plot was obtained by sampling 500 intervals from a 5000-node pop/sim causet.  
The repertoire of possible (\emph{Volume}($I$), \emph{OrderingFraction}($I$)) pairs is very limited here, 
with volumes below 6 nodes, and the single value 1 for the ordering fraction.   
The preferential attachment policy of the pop/sim algorithm 
-- new nodes prefer to connect with the popular nodes with lowest birth times -- 
yields causets with very short chains, thus small intervals, that are totally ordered 
(a necessary and sufficient condition for the ordering fraction to be 1) 
or almost totally ordered.  
An exhaustive search of all the intervals from this specific 5000-node causet reveals that 
the lowest possible ordering fraction value is indeed 21/22, achieved by only three intervals of volume 12. 

\begin{figure}[h]
\centering
\includegraphics[scale=.6]{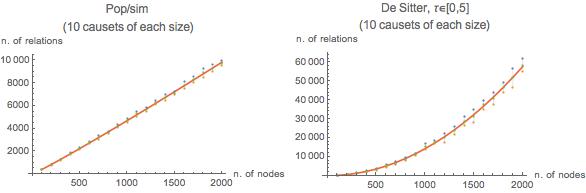}
\caption{{\footnotesize Growths of the number of relations (edges) in pop/sim and de Sitter causets as a function of causet size,
ranging from 100 to 2000 nodes, in steps of 100.  For each size we compute the number of relations for 10 causets, and plot
the mean and standard deviation.  These data are matched against their linear and quadratic fitting functions.}}
\label{fig:fullPopSimAndFullDeSitterEdgeGrowths}     
\end{figure}

In light of the fact, claimed in \cite{ref:Krioukov2012}, that pop/sim and sprinkled deSitter causets 
share the same asymptotic dynamics, we would be interested in comparing their ordering fraction spectra, 
shown, respectively, in the upper-right and lower-right plot of 
Figure \ref{fig:ordFractSpectraForCausetClasses}.
Unfortunately, the  limited interval volumes that we can attain for pop/sim causets makes the comparison unfeasible.  
However, much lower ordering fraction values are achieved for \emph{whole} pop/sim causets, rather than for their intervals.  
In particular, we experimentally find that the ordering fraction of a $k$-node pop/sim causet is \emph{not} constant: 
it decreases with $k$, and is an $O(k^{-1})$ function.  
Since the ordering fraction is 
$R(k)/{k \choose 2}$ and 
${k \choose 2}$ is $O(k^{2})$, 
this result indicates that $R(k)$ -- the number of edges in the transitively closed causet -- is $O(k)$, 
like the number $RR(k)$ of edges in the original, 'raw' causet.  
%
This linear growth is illustrated in Figure \ref{fig:fullPopSimAndFullDeSitterEdgeGrowths}-left.
%
%
%
(Note that the linearity of $RR(k)$ is a direct consequence of adding a fixed number of edges with each new node.) 

On the other hand, we also find experimentally that the growth rate of the edge count for \emph{whole} $k$-node
de Sitter sections is $O(k^{2})$ 
(Figure \ref{fig:fullPopSimAndFullDeSitterEdgeGrowths}-right),
like that indirectly revealed by Figure \ref{fig:ordFractSpectraForCausetClasses} above for $k$-node de Sitter \emph{intervals}.
Thus, in spite of the asymptotic similarity of full de Sitter causets and pop/sim causets relative to node degree distributions, 
these causet classes differ in the growth rates of their edges, which we found to be, respectively, 
$O(k^{2})$ and $O(k)$. 


The conclusion we can draw from the inspection of the above plots is that ordering fraction spectra, 
while useful for possibly detecting causet dimensionality (limited to sprinkled causets from flat spacetime, 
according to \cite{ref:MeyerThesis1989}), 
are not relevant for revealing causet Lorentzianity, since, again, 
they do not separate sprinked causets (Lorentzian) from a regular grid (non-Lorentzian).

%
%
\section{Counting links} 
\label{sect:CountingLinks}

In this section we try to characterise causets by collecting statistical information on 
edge counts and paths for their \emph{transitively reduced} forms.  
'Links' is the name commonly adopted for indicating the essential edges left in the causet after transitive reduction.  
(We shall only deal with finite causets, thus finite partial order relations.
Recall that the transitive reduction of a finite relation $R$ is the smallest relation that admits the same transitive closure of $R$; 
when $R$ is acyclic, its transitive reduction is unique.) 

A first simple indicator is suggested by this quote from D. Rideout  \cite{ref:RideoutHomePage}:

\begin{quotation}
"\emph{The 'usual' discrete structures which we encounter, e.g. as discrete approximations to spatial geometry, 
have a 'mean valence' of order 1. e.g. each `node' of a Cartesian lattice in three dimensions has six nearest neighbors. 
Random spatial lattices, such as a Voronoi complex, will similarly have valences of order 1 [...].  
Such discrete structures cannot hope to capture the noncompact Lorentz symmetry of spacetime. 
Causal sets, however, have a `mean valence' which grows with some finite power of the number of elements in the causet set.  
It is this 'hyper-connectivity' that allows them to maintain Lorentz invariance in the presence of discreteness.}"

\end{quotation}%

It should perhaps be clarified that we cannot meaningfully apply the concept of Lorentz invariance directly to a causet $C$, 
simply because there are no coordinates to be Lorentz-transformed.  
But when $C$ comes already embedded in some manifold $M$, with its coordinate system, 
we may keep the information on node coordinates, apply the transformation to $M$, 
and see its effect on $C$, which is now dragged and reshaped by the operation.  
When $C$ is obtained by a Poisson sprinkling in $M$ it can be shown that in the equivalence class 
of all Lorentz-transformed embedded versions of $C$ there is \emph{no preferred element} 
that we can pick out, while if we embed, for example, a regular grid, this is not the case.  
\footnote{
The sprinkled causet vs. grid (or 'diamond lattice') example is often mentioned in the literature, 
e.g. in \cite{ref:SorkinValdivia2003} and \cite{ref:Krioukov2012}.
Of course, a typical, random-looking sprinkled causet gets Lorentz-transformed into another
random-like causet which is, strictly speaking, different from the original.
Their indistinguishability is to be intended in statistical sense.
For example, the density of points, or
the expected distance and angle distribution from a generic point to its closest neighbor are unaffected by the transformation.
Under a statistical mechanics metaphor, they are two equivalent micro-states yielding the same perceived macro-state.
The case of the regular 'diamond lattice' is clearly different, due to the macroscopic 'polarisation' induced in it by the Lorentz transformation.
}
In the above quote, Rideout establishes a connection between this invariance property of sprinkled causets 
(embeddable by definition) and the growing degree of their nodes.

Note that this degree refers, here and in the rest of the section, to the links, as defined above.  
Our general approach is to look for statistical properties that are necessarily satisfied 
by causets whose embeddability is guaranteed, namely sprinkled causets, 
and then use these properties for checking directly the given $C$, without worrying about its embeddability.  
As a first property, we look at the growth rate of node degrees in transitively reduced sprinkled causets, 
keeping in mind that the first concern will be to detect an actual growth of node degrees with graph size, 
i.e.\ to exclude the above mentioned $O(1)$ growth.

%
\subsection{Growth rate of node degrees}
\label{subsect:GrowthRateOfNodeDegrees}
%

In \cite{ref:bombelliReply88} Bombelli et al.\ mention that, considering the causet $C[s, t]$ 
obtained from uniformly sprinkling points in an order interval $I[s, t]$ of height $T$ of d-dimensional Minkowski space 
($T$ being the Lorentz distance between $s$ and $t$), the number of nearest neighbors of root $s$ in the interval 
-- the number of outgoing links -- 
grows like  $Log(T)$ for $d =2$, and like $T^{d-2}$ for $d \geq 3$, provided that the sprinkling density is kept constant.  
Equivalently, we can consider an interval of fixed, unitary height 
and increase the number $k$ of nodes sprinkled in it (thus, the density), 
in which case the root degree $deg(k)$ has growth $O(Log(k^{1/2})) = O(Log(k))$ for $d = 2$, 
and $O(k^{((d-2)/d))}$ for $d \geq 3$.  
\footnote{We detect a problem in \cite{ref:EichhornMizera2013}, equation (8), which indicates that the number of links from the source of a 
$(d+1)$-dimensional interval with $N$ nodes is proportional to $N^{(d-1)/2}$ for $d > 1$.  
In light of the above results,  the formula should read $N^{(d-1)/(d+1)}$.
}

In this subsection we carry out explicitly the analysis of cases $d = 2$ and $d = 3$ 
(which is not given in \cite{ref:bombelliReply88}),
obtain accurate asymptotic expressions for $deg(k)$, not just their growth rate, 
and verify their agreement with data from simulations.

\subsubsection*{2D sprinkled Minkowski intervals}

\begin{figure}[h]
\centering
\includegraphics[scale=.59]{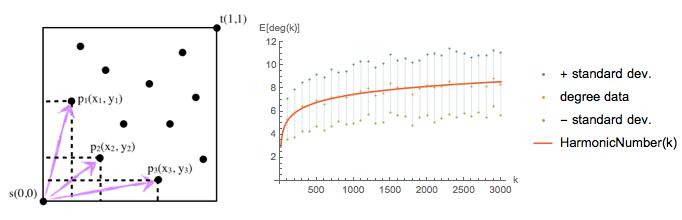}
\caption{{\footnotesize 
Left: sprinkling 11 points in the unit box and finding the three links from the root.  
Right: $O(log(k))$ growth of the expected root degree of a 2D Minkowski causet interval.  
For each value of $k$, ranging from 100 to 3000 in steps of 100, the expected degree was obtained by
averaging over 100 sprinklings in the unit box, each consisting of $k$ points. 
Standard deviations are show, as well as the match between averaged data and analytical expression.}}
\label{fig:rootLinkGrowthInSprink2Dv8}     
\end{figure}

For the analysis of the two-dimensional case we find it convenient, as done before, 
to represent the sprinkled interval as a unit square $[0, 1]\times[0, 1]$ in Euclidean 2D space, 
where $s(0, 0)$ is the interval root;
$k$ points are uniformly distributed inside the square 
(see Figure \ref{fig:rootLinkGrowthInSprink2Dv8}, where $k = 11$ - source and sink are excluded from the count).  

%

By the definition of 'link', $deg(k)$ corresponds now to the number of points, out of $k$,
that form empty rectangles $R_{i}(s,p_{i})$; 
the latter are identified by their lower-left and upper-right vertices.  
Three such rectangles are depicted in Figure \ref{fig: rootLinkGrowthInSprink2Dv8}-left, with dotted lines; 
their purple diagonals are the only links from the root $s$, for the causet (not shown) associated with the sprinkling.

Finding the expected value $E[deg(k)]$ of random variable $deg(k)$ is equivalent to 
finding the expected number of so called 'Pareto-optimal' elements for the same set of points.
\footnote{
http://math.stackexchange.com/questions/206866/expected-number-of-pareto-optimal-points.}
In brief, the analysis is as follows.
The probability of point $p_{i}$ to be the endpoint of a link $s \rightarrow p_{i}$, under the condition that its coordinates are $(x_{i}, y_{i})$,
depends on these coordinates, and corresponds to the probability of all other $k-1$ points 
to fall \emph{outside} rectangle $R(s, p_{i})$:
\begin{equation}
Prob(s \rightarrow p_{i} \textnormal{ is a link } | \; p_{i}\textnormal{'s coordinates are } (x, y)) = (1-xy)^{k-1},
\label{eq:condLinkProb2D}
\end{equation}
since each of those points has probability $xy$ to fall \emph{inside} the rectangle.
For finding the unconditioned probability of $p_{i}$ yielding a link, we let its coordinates range over the unit square:
\begin{equation}
Prob(s \rightarrow p_{i} \textnormal{ is a link}) = \int_{0}^{1} \int_{0}^{1} (1-xy)^{k-1} dx dy = H(k)/k,
\label{eq:linkProb2D}
\end{equation}
where $H(k)$ is the Harmonic Number function: $H(k) = \sum_{i=1}^{k}\frac{1}{i}$.
The expected number of links is then obtained by adding the contributions of all $k$ points:
\begin{equation}
E[deg(k)] = H(k).
\label{eq:expectedMeanDegFor2Dsprink}
\end{equation}

Figure \ref{fig: rootLinkGrowthInSprink2Dv8}-right is a plot of experimental data for the root degree of sprinkled, 
transitively reduced 2D Minkowski interval causets, for densities from $k = 100$ to $k = 3000$, in steps of 100 (the dots).
Each data point was obtained by averaging over 100 intervals of fixed density. 
These data are
matched by the $O(log(k))$ theoretical expected degree $E[deg(k)]$ of equation \ref{eq:expectedMeanDegFor2Dsprink} (solid line).  
Recall that $H(k) \in O(log(k))$  since it asymptotically approaches $log(k) + \gamma$, where $\gamma$ is the
Euler-Mascheroni constant 0.5772.

\subsubsection*{3D sprinkled Minkowski intervals}
%

\begin{figure}[h]
\centering
\includegraphics[scale=.4]{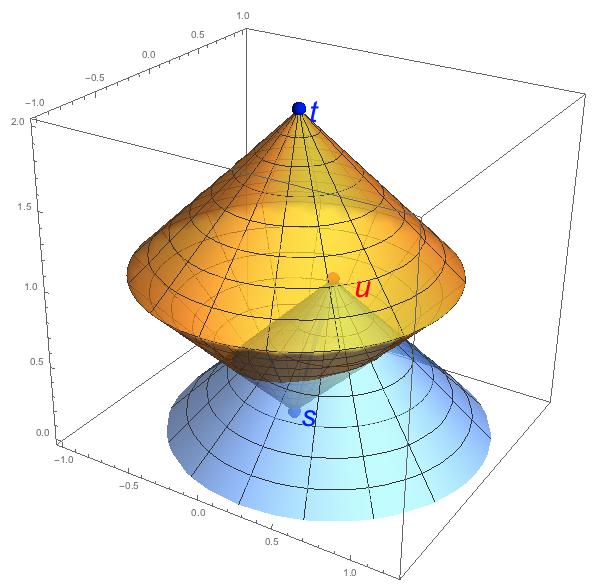}
\caption{{\footnotesize Interval $I[s, t]$ in Minkowski space $M^{(1,2)}$ (in yellow), 
and a sub-interval $X[s, u]$, for a point $u$ inside $I[s, t]$.  
The analysis of the degree of $s$ makes use of the volume of $X[s, u]$ as a function of the position of $u$.}}
\label{fig:biconeAndCone}     
\end{figure}

For the analysis of the 3D case we use a similar approach.  
We sprinkle a set $S$ of $k$ points in a 3D interval $I[s, t]$ of Minkowski space $M^{(1,2)}$, 
assuming, w.l.o.g.,  $s = (0, 0, 0)$ and $t = (0, 0, 2)$; this is depicted as the yellow double cone in  Figure \ref{fig:biconeAndCone}.

Any point  $u(r, \theta, z)$ - using cylindrical coordinates -  inside $I[s, t]$ identifies a sub-interval $X[s, u]$, delimited, in Figure \ref{fig:biconeAndCone}, 
by the lower yellow cone and the blue cone. 
The probability of a sprinkled point to fall \emph{inside} $X[s, u]$ is $vol(X[s, u]) / vol(I[s, t])$, where
$vol(I[s, t]) = 2\pi/3$.  
For symmetry, the volume of the 'skew' interval $X[s, u]$ depends only on $u$'s coordinates $r$ and $z$,
and can be calculated to be: 
\footnote{A quick and elegant way to carry out this calculation was suggested by an anonymous referee.
The element $dV$ of volume in Minkowski space is Lorentz invariant, hence
$vol(X[s, u])$ must be a function of $(z^{2} - r^{2})$. Furthermore 3D volume must scale as the cube
of the linear measure, hence it is proportional to $(z^{2} - r^{2})^{3/2}$.  The multiplicative constant $\pi/12$
is obtained by explicitly computing a specific volume, e.g. for $z=2$ and $r=0$.
}
\begin{equation} 
vol(X[s, u]) = \frac{\pi}{12}(z^2-r^2)^{3/2}.
\label{eq:skewBiconeVol}
\end{equation}
Then, analogous to the 2D case (equation (\ref{eq:condLinkProb2D})), 
we obtain the conditional probability of $u$ to top an empty subinterval, thus yielding a link:
\begin{equation} 
Prob(s \rightarrow u \textnormal{ is a link } | \; u\textnormal{'s coordinates are } (r, \theta, z)) = (1-\frac{1}{8}(z^2-r^2)^{3/2})^{k-1}.
\label{eq:condLinkProb3D}
\end{equation}

Let $P_{link}(r,z)$ concisely denote the above conditional probability.
Analogous to equation \ref{eq:linkProb2D},
the probability of the \emph{generic} point $u$ to be the endpoint of a link $s \rightarrow u$ is given by a triple integral in cylindrical coordinates, divided by the volume of the region $I[s,t]$ of integration:
\begin{equation} 
Prob(s \rightarrow u \textnormal{ is a link}) = \frac{\int_{I[s, t]} r P_{link}(r,z) dr \, d\theta \, dz}{\frac{\pi}{12}(z^2-r^2)^{3/2}}.
\label{eq:linkProb3D}
\end{equation}

The expected number of links will then be obtained, again, by adding the contributions of all $k$ points:
\begin{equation} 
E[deg(k)] = k \; Prob(s \rightarrow u \textnormal{ is a link}).
\label{eq:ExpectedDegFromProb3D}
\end{equation}

For the calculations it is now convenient to distinguish between  \emph{lower cone} and \emph{upper cone}, 
corresponding to $z \in [0,1]$ and $z \in [1,2]$, and express the above probability as follows:
\begin{eqnarray} 
Prob(s \rightarrow u \textnormal{ is a link}) =  \frac{1}{2}Prob(s \rightarrow u \textnormal{ is a link } | \; u \in \textnormal{ lower cone})  \nonumber \\ 
+ \frac{1}{2}Prob(s \rightarrow u \textnormal{ is a link } | \; u \in \textnormal{ upper cone}),
\label{eq:probSplit}
\end{eqnarray}

For the lower cone, of volume $\pi/3$, we obtain:
\begin{eqnarray} 
Prob(s \rightarrow u \textnormal{ is a link } | \; u \in \textnormal{ lower cone}) =  \nonumber
\frac{\int_{z=0}^{1} \int_{\theta=0}^{2\pi} \int_{r=0}^{z}  r P_{link}(r,z) dr \, d\theta \, dz}{\pi/3} \\ 
= \frac{2^{4-3k}7^{k}-16}{k} + 3 \; {}_{2}F_{1}(\frac{2}{3}, 1-k, \frac{5}{3}, \frac{1}{8}),
\label{eq:linkProb3Dlower}
\end{eqnarray}
where ${}_{2}F_{1}$ is the Gaussian, ordinary Hypergeometric function.
 
\begin{figure}[h]
\centering
\includegraphics[scale=.54]{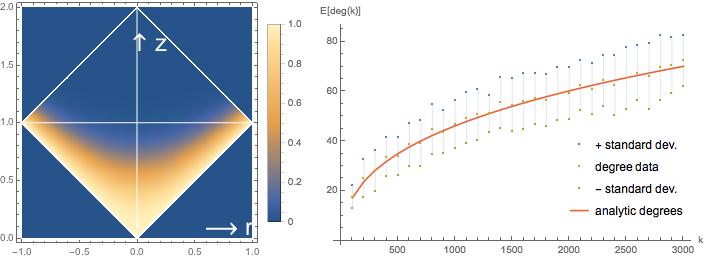}
\caption{{\footnotesize 
Left: Density plot for the probability of a point $u(r, \theta, z)$ to be the endpoint of a link $s \rightarrow u$,
as a function of $r$ and $z$: the contribution from the upper cone is almost null.
Right: Growth of the expected degree $E[deg(k)]$ of the root node of a causet 
obtained by sprinkling $k$ points in 3D Minkowski intervals;
match between experimental data for $E[deg(k)]$ and the
analytic expression of  eq. (\ref{eq:ExpectedDegFromProb3Dexplicit}). 
Data points refer to values of $k$ from 100 to 3000, in steps of 100. 
Each point was obtained by averaging over 50 intervals of fixed density $k$.}}
\label{fig:linkDegGrowthInMink3d}     
\end{figure}

Integration for the upper cone is harder, however we can safely ignore this component in light of the fact 
that the integrand function $P_{link}(r,z)$ basically vanishes in the upper cone, 
as illustrated by the density plot in Figure \ref{fig:linkDegGrowthInMink3d}-left.

%
%
In conclusion, in light of equations (\ref{eq:ExpectedDegFromProb3D})-(\ref{eq:linkProb3Dlower}) we 
derive the following asymptotic expression for the root degree as a function of $k$:
\begin{equation} 
E[deg(k)] \approx 8((7/8)^{k} - 1) + (3/2) k* {}_{2}F_{1}(\frac{2}{3}, 1-k, \frac{5}{3}, \frac{1}{8}).
\label{eq:ExpectedDegFromProb3Dexplicit}
\end{equation}
In Figure \ref{fig:linkDegGrowthInMink3d}-right we show the match between 
the analytic expression of eq.\ (\ref{eq:ExpectedDegFromProb3Dexplicit}) and experimental data.

In spite of the many thousand simplified forms for special cases of the Hypergeometric function
\footnote{See, for example, http://functions.wolfram.com/HypergeometricFunctions/Hypergeometric2F1/03/
which lists 111,271 formulas.}, 
we could not find a simplified expression for the specific parameter setting of eq.  (\ref{eq:ExpectedDegFromProb3Dexplicit}),
from which to explicitly detect the $O(k^{1/3})$ growth rate of the expected node degree, as mentioned in \cite{ref:bombelliReply88}.
However, an accurate verification of this growth rate for the function at the r.h.s.\ of eq.\ (\ref{eq:ExpectedDegFromProb3Dexplicit})
is readily obtained in \emph{Mathematica}, 
by using function fitting facilities and the available implementation of the Hypergeometric function
(details are omitted).

On the other hand, an explicit derivation of the $O(k^{1/3})$ result is achieved by approximating
the probability in eq. (\ref{eq:condLinkProb3D}) by its negative exponential Poisson form.

The Poisson probability to find $h$ points inside interval $X[s, u]$, out of  $k-1$ points sprinkled inside $I[s, t]$, is 
$ \frac{\lambda^h}{h!} e^{-\lambda}$, where $\lambda$ is the expected value of $h$: 
\begin{equation} 
\lambda = E[h] = (k-1)\frac{vol(X[s,u])}{vol(I[s,t])} = (k-1)\frac{1}{8}(z^2-r^2)^{3/2},
\end{equation}
and the probability to find \emph{zero} points  inside $X[s, u]$ reduces to $e^{-\lambda}$.  
This yields a revised expression for the conditional probability of (\ref{eq:condLinkProb3D}):
\begin{equation} 
Prob(s \rightarrow u \textnormal{ is a link } | \; u\textnormal{'s coordinates are } (r, \theta, z)) = e^{-\frac{1}{8}(k-1)(z^2-r^2)^{3/2}}.
\label{eq:condLinkProb3Dpoisson}
\end{equation}

Let $P'_{link}(r,z)$ denote the above conditional probability.
As before, the unconditioned probability of a generic point to yield a link is obtained by integration, 
and, again, we restrict to the lower cone. 
\begin{eqnarray} 
Prob(s \rightarrow u \textnormal{ is a link } | \; u \in \textnormal{ lower cone}) =  \nonumber
\frac{\int_{z=0}^{1} \int_{\theta=0}^{2\pi} \int_{r=0}^{z}  r P'_{link}(r,z) dr \, d\theta \, dz}{\pi/3} \\ 
= -\frac{2(8 - 8 \; e^{\frac{1-k}{8}}+(k-1)E_{1/3}(\frac{k-1}{8}))} {k-1} + \frac{12 \;\Gamma(5/3)}{(k-1)^{2/3}},
\label{eq:linkProb3DlowerPoisson}
\end{eqnarray}
where $E_{n}(x)$ is the exponential integral function.  
Again we obtain an approximation of the expected degree by multiplying the above probability by $k/2$,
which can be thought of also as the expected number of points that fall in the lower cone. A little manipulation yields:
\begin{equation} 
E[deg(k)] \approx k ( \frac{8(e^{\frac{1-k}{8}}-1)}{k-1} - E_{1/3}(\frac{k-1}{8}) + \frac{6 \; \Gamma[5/3]}{(k-1)^{2/3}}).
\label{eq:ExpectedDegFromProb3DexplicitPoisson}
\end{equation}

The first and second terms converge, respectively, to -8 and 0.  
The leading, third term explicitly reveals the $O(k^{1/3})$ growth,
confirming, for dimension $d=3$, the $O(k^{\frac{d-2}{d}})$ growth rate of node degrees mentioned in \cite{ref:bombelliReply88}.
(Integration relative to the upper cone yields slightly more complicated terms, but one finds that
they all vanish as $k \rightarrow \infty$.) 

How about the other causet classes?

\subsubsection*{2D sprinkled de Sitter intervals}
%

Our analysis of the the expected node degree growth for sprinkled de Sitter intervals is limited to the experimental approach,
and to the sole 2D case. 

We considered a fixed interval $I[s, t]$ of the hyperboloid representing 2D de Sitter spacetime
(see subsection \ref{subsect:DeSitterSprinkling}), with time parameter $\tau$ ranging between
$\tau_{min} = 0$ at $s$ to $tau_{max} = 5$ at $t$.
Then, for each considered value $k$ ($k$ ranging from 100 to 2000 in steps of 100)
we built 100 directed graphs, each obtained by sprinkling $k$ points in $I[s, t]$, finding causal edges,
and applying transitive reduction; the latter was the bottleneck of the whole computation.
The expected root degree $E[deg(k)]$ was obtained, for each $k$, by averaging over the 100 samples.

\begin{figure}[h]
\centering
\includegraphics[scale=.54]{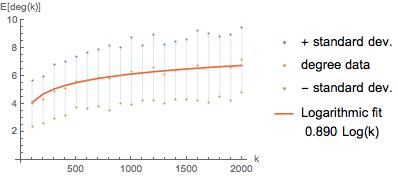}
\caption{{\footnotesize 
Growth of the expected degree $E[deg(k)]$ of the root of a causet
obtained by sprinkling $k$ points in a 2D de Sitter interval spanning time interval $[0, 5]$;
logarithmic fitting. 
Data points refer to values of $k$ from 100 to 3000, in steps of 100. 
Each point was obtained by averaging over 100 causets with a fixed value of $k$. Standard deviations are shown.}}
\label{fig:linkDegGrowthInDeSitter2D}     
\end{figure}

The expected degree growth appears to be $O(Log(k))$ as for the Minkowski case, as illustrated in Figure \ref{fig:linkDegGrowthInDeSitter2D}.  

%

In fact, there is no reason to expect sprinkled de Sitter causets to manifest a behaviour qualitatively different from 
that of sprinkled Minkowski causets, in the sense that we can still exclude the extreme $O(1)$ scenario.  
An informal justification for this claim is obtained by considering the ultimate reason why 
node degrees steadily grow in Minkowski sprinkled causets: 
as the sprinkling density increases, 
new nodes $x$ will indefinitely appear that are close enough to the future lightcone of the interval root $s$ to create a link 
$s \rightarrow x$ that contributes to the growth of the degree of $s$.  
This argument should not be affected when introducing curvature.

\subsubsection*{Percolation causet intervals}

The node degree growth scenario for percolation causet intervals is quite different,
and is ultimately determined by the peculiar phenomenon of \emph{posts}.
A \emph{post} in a partial order is an element that happens to be related with \emph{all} other elements.
In other words, a post $x$ in a causet creates a bipartition of the whole set of nodes into the future and the past lightcone of $x$.  
In \cite{ref:AlonEtAl94} it is established that, when the edge probability $p$ is constant,
almost surely posts keep appearing indefinitely, as the number $n$ of elements grows. 
\footnote
{If the edge probability $p$ is allowed to vary as a function of $n$,
then  \cite{ref:BolBri97a} proves that almost surely posts arise when  $n p^{-1} e^{\frac{-\pi ^{2}}{3p}} \rightarrow \infty$,
as $n \rightarrow \infty$, while they almost surely do not arise when this quantity tends to 0. 
In this paper we only consider the case of a constant $p$, which implies that the first case applies.
}
This yields a  'bouncing universe' picture where each post in the sequence $p_{1}, p_{2}, ... p_{n}, ... $
represents a Big Crunch/Big Bang event.  

Let $C$ be a percolation causet built by using a fixed edge probability $p$, and
let $C_{i}$ denote the interval $I[p_{i}, p_{i+1}]$ between two adjacent posts.
An interval $I[x, y]$ of $C$ of growing size $n$ will eventually span across multiple posts, i.e.
$x$ and $y$ will fall in separate $C_{i}$'s. 
The expected degree of $x$ will be bounded by the expected size of the $C_{i}$ where it falls,
which does not depend on $n$.  Thus, the growth of the expected degree of $x$ as a function of
$n$ is in $O(1)$.

Note that this (null) growth rate is consistent with the asymptotic dimension 1
suggested by the ordering fraction spectrum 
in the lower-central plot of Figure \ref{fig:ordFractSpectraForCausetClasses}.

\begin{figure}[h]
\centering
\includegraphics[scale=.54]{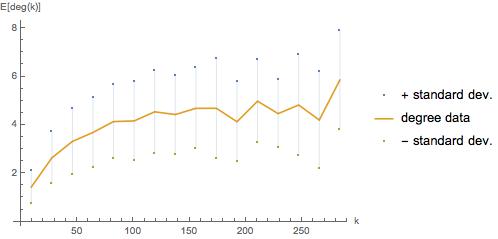}
\caption{{\footnotesize 
Histogram of the expected root degree $E[deg(k)]$ for percolation intervals of various sizes. 
10,000 intervals have been used, each picked at random from a different 1000-node percolation causet with edge probability $p = 0.01$.
The intervals were grouped into 15 slots of equal width
according to their sizes (number of nodes). 
For each slot, the mean root degree of the corresponding intervals is shown, with standard deviations.}}
\label{fig:rootLinkDegGrowthFor10kPercolInts}     
\end{figure}

Figure \ref{fig:rootLinkDegGrowthFor10kPercolInts} plots data for the root degree of percolation causet intervals.  
%
%
This plot is provided mainly for uniformity with our treatment of the other causet cases,
but the data we could collect by simulation does not provide a clear indication of the anticipated $O(1)$ behaviour.
This is not surprising since the $O(1)$ result crucially depends on the existence and inter-distances of posts, and the occurrence of these
special nodes is extremely rare.

To get a feel for their rarity we have collected some statistical data on their frequency, following two alternative criteria.

\begin{figure}[h]
\centering
\includegraphics[scale=.5]{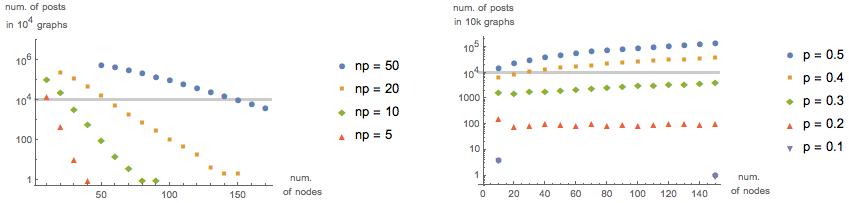}
\caption{{\footnotesize 
Left: Number of posts cumulatively found in 10,000 $n$-node percolation causets for various values of $n$.  
The edge probability is $p = h/n$
for $h = 5, 10, 20$.
Right: Number of posts cumulatively found in 10,000 $n$-node percolation causets for various values of $n$,
and for edge probabilities $p$ from 0.1 to 0.5 in steps of 0.1.  
In both plots
the points below the  horizontal line at $10^4$ correspond to settings of the $n$ and $p$ parameters
for which the causet is expected to include less than 1 post.}}
\label{fig:rarityOfPosts}     
\end{figure}

Figure \ref{fig:rarityOfPosts}-left shows the cumulative number of posts achieved by 10,000 $n$-node percolation causets for various values of $n$.
The edge probability $p$ here depends on $n$, and is defined as $h/n$, where the four cases  $h = 5, 10, 20, 50$ are considered.  
It is trivial to see that this choice of variable edge probability keeps the expected
node degree (in-degree + out-degree) constant.  
Note that we include in the count of posts the degenerate cases of node 1 and node $n$, whenever the
degree of these nodes in the transitively closed graph happens to be $n-1$.

In Figure \ref{fig:rarityOfPosts}-right we deal with the different scenario of a constant edge probability $p$,
and consider the cases $p = 0.5, 0.4, 0.3, 0.2, 0.1$.  In the limit case of $p = 1$ the order becomes total and all nodes are posts.
But as $p$ decreases, post occurrences are dramatically reduced.
When $p = 0.1$, for most values of $n$ the number of posts found in 10,000 causets drops to zero, failing to be represented
in the \emph{Log} plot.

Both plots include a horizontal line at ordinate $10^4$:
since we use $10^4$ raw percolation causets for each $n$,
any point in the plot that falls below the line corresponds to
causets for which the expected number of posts is less than 1. 

Let us now go back to the histogram of Figure \ref{fig:ordFractSpectraForCausetClasses}.
The causet intervals considered in that figure were derived from 1000-node percolation
causets with edge probability $p = 0.01$, which yields $np = 10$:  
for this parameter setting, the diagram of Figure \ref{fig:rarityOfPosts}-left clearly indicates that the chances to find
even a single post in a 1000-node causet are vanishingly small, and we conclude that posts are not involved
in the shaping of that histogram.

\subsubsection*{Pop/sim causet intervals}

For pop/sim causets (see Subsection \ref{subsect:PopularitySimilarityDynamics}), 
the expected degree (counting outgoing links) of the root $s$ of a $k$-node \emph{interval} $I[s, t]$ does \emph{not} grow with $k$,
i.e. it is an $O(1)$ function, regardless of the value of the parameter $m$ representing 
the number of new edges $x_{1}\rightarrow n$ ... $x_{m}\rightarrow n$
contributed by each new node $n$ to the growing raw causet. 
The reason is as follows.

Consider an $i$-indexed family of \emph{independent}, pop/sim $k$-node intervals $I_{i}[s_{i}, t_{i}]$,
with $i$ ranging in some (large) index set $M$, and
where, for independence, each interval is extracted from a separate raw pop/sim causet $G_{i}$, $i \in M$.
For a generic \emph{raw} pop/sim DAG $G(E, N)$ with fixed in-degree $m$, each newly added node $x$
contributes \emph{exactly} $m$ edges - those that reach the node -, 
except for the initial $m$ nodes, that have no incoming edge. 
Thus $|E| < m|N|$, which implies, for out-degrees, that
$Mean\{outDegree(x) | x \in N\}< m$.  
This expectation must also be valid for the set of interval roots $s_{i}$:
$Mean\{outDegree(s_{i}) | i \in M\}< m$.  
Each $outDegree(s_{i})$ is computed relative to the \emph{complete} corresponding \emph{raw} graph $G_{i}$.
Letting $outDegree'(s_{i})$ and $outDegree"(s_{i})$ denote the out-degree of $s_{i}$ relative, respectively, 
to interval $I_{i}[s_{i}, t_{i}]$ and to the transitive reduction of the latter (which still must have $k$ nodes), 
we have:
$outDegree(s_{i}) \geq outDegree'(s_{i}) \geq outDegree"(s_{i})$.  We are using here the obvious fact that 
the \emph{links} are a subset of the \emph{raw} edges.
This allows us to conclude, restricting to intervals and links, that 
$Mean\{outDegree"(s_{i}) | i \in M\}< m$.
Then, the fact that the above inequalities do not depend on $k$ establishes the $O$(1) growth result.

\subsubsection*{Regular grid interval}

The case of regular grids trivially falls into the $O$(1) growth scenario.  
As pointed out in the above quote \cite{ref:RideoutHomePage}, in these graphs (or even in their randomized variants) 
node degrees do not grow with the number of nodes.

\subsubsection*{Assessing the node degree growth indicator: the case of the \emph{irrational grid}}
\label{subsubsect:irrationalGrid}

Looking at node degree growth is a useful way to investigate causet Lorentzianity.
In particular, avoiding an $O(1)$ growth for interval root degrees is a \emph{necessary} condition 
for achieving the hyper-connectivity of Lorentzian, sprinkled causets.
This allows us to rule out percolation causets, grid causets, \emph{and} pop/sim causets.
 
Furthermore, once an unbounded growth rate is detected, a more accurate characterisation of the growth
- whether logarithmic or polynomial of certain degree -
might provide us with a dimensionality estimate for the causet under study,
based on the well characterised growth rates for sprinkled causets. 

Then, in order to come up with some reasonable definition of Lorentzianity
in the discrete setting
we might be tempted to promote unbounded node degree growth to the status of a necessary \emph{and} sufficient condition.
For this to be a good choice, 
any transitively reduced causet  manifesting an unbounded node degree growth 
should appear to us as a 'good' Lorentizan causet.
The following counter-example suggests that this is not the case. 

Consider a conceptually simple variant of the square grid, that we call 'irrational grid', in which
the points in 2D Minkowski spacetime $M^{(1,1)}$ have coordinates $(x\sqrt{2}, t)$, with integers $x$ and $t$ ranging in $(-\infty, +\infty)$.
It is easy to establish that the degree of any node in the transitively reduced causet associated with these points is infinite.

\begin{figure}[h]
\centering
\includegraphics[scale=.24]{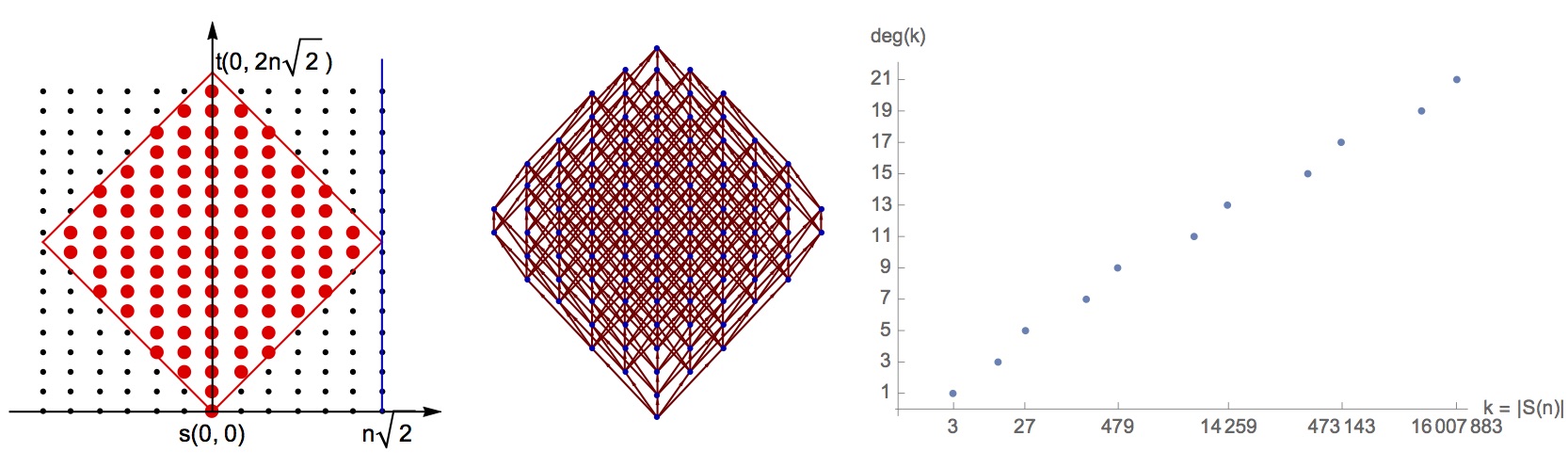}
\caption{{\footnotesize 
Left: irrational grid and diamond for n = 5. 
Center: transitively reduced causet for the diamond points. Root degree is 5. 
Right:  logarithmic growth of the root degree as a function of diamond size (log-linear plot)}}
\label{fig:IrrationalGridNodeDegreeGrowth}     
\end{figure}

More precisely (see Figure \ref{fig:IrrationalGridNodeDegreeGrowth}-left)
if $I[s, t]$ is a diamond interval in $M^{(1,1)}$ between $s(0,0)$ (a grid point) and $t(0, 2n\sqrt{2})$ (not a grid point), 
where positive integer $n$ is the 
index of the column of grid points delimiting the diamond at the r.h.s., 
and $S(n)$ is the set of irrational grid points that fall inside $I[s, t]$,
then we can readily estimate the growth of $|S(n)|$ to be, asymptotically, 
 $|S(n)| \approx  2n^{2}\sqrt{2}.$ 
 %
 
On the other hand, letting 
$C(n)$ denote the transitively reduced causet with root $s$ derived from the points of set $S(n)$
(Figure \ref{fig:IrrationalGridNodeDegreeGrowth}-center),
and 
$deg(n)$ denote the degree of $s$ in $C(n)$, 
we find that the precise initial values of $|S(n)|$, and the values of $n$ 
that mark an increment of $deg(n)$, are as listed in Table \ref{tab:irrationalGridNumbers} below. 

\begin{table}[ht]
\begin{center}
\begin{footnotesize}
 \begin{tabular}{||c || c | c | c c c c c | c c c c c | c c c ||} 
 \hline
 \emph{n} & 1 & 2 & 3 & 4 & 5 & 6 & 7 & 8 & 9 & 10 & 11 & 12 & 13 & 14 & ... \\ 
 \hline 
 $|S(n)|$ & \bf{3} & \bf{12} & \bf{27} &46 &71 &101 &138 & \bf{181} &230 &283 &342 &408 & \bf{479} &556 & ... \\
 \hline
 \emph{deg(n)} &\bf{1} &\bf{3}   &\bf{5} &5 &5 &5 &5  &\bf{7} &7 &7 &7 &7   &\bf{9} &9 &... \\
  \hline
\end{tabular}
\caption{
{\footnotesize Data for the irrational grid.  
First row: Semi-width of diamond interval (/$\sqrt{2})$.  
Second row: number of points falling inside the diamond, and size of causet $C(n)$.
Third row:  degree of causet root.}
}
\label{tab:irrationalGridNumbers}
\end{footnotesize}
\end{center}
\end{table}

Degree values, as shown in the third row, are always odd and grow by +2 increments, for the symmetry of the graph.  
Furthermore, the analysis of the lenghts $\ell_{i}$ of the 'runs' of equal values in the degree sequence reveals that $\ell_{i} = \ell_{i+1}$ for all odd $i$'s - 
e.g. $\ell_{1} = \ell_{2} = 1$, or $\ell_{3} = \ell_{4} = 5$ - and that the ratio $\ell_{i} / \ell_{i-1}$, for $i \geq 3$ and odd, 
quickly converges to constant value $3 + 2\sqrt{2} = 5.82843$, which implies a logarithmic node degree growth.
\footnote{We thank an anonymous referee for suggesting the simple expression $3 + 2\sqrt{2}$ for constant 5.82843.
The fact that $\sqrt{2}$ appears in it is perhaps not surprising, given the construction of Figure \ref{fig:IrrationalGridNodeDegreeGrowth}-left.}


This growth is illustrated in Figure \ref{fig:IrrationalGridNodeDegreeGrowth}-right, which plots the $(|S(n)|, deg(n))$ pairs 
listed in Table \ref{tab:irrationalGridNumbers} and beyond.  
It turns out that this degree growth and the corresponding growth for sprinkled 2D causets 
(Figure \ref{fig:rootLinkGrowthInSprink2Dv8}-right)
are not only analogous in their logarithmic character, but also very close numerically.
In this respect, the irrational grid could be seen as an adequate \emph{regular} counterpart of the \emph{random}-looking, sprinkled 2D causet.

In summary, while the node degree growth indicator correctly induces us to rule out 
the causet obtained from the original, square grid (or from any \emph{rational} grid, for that matter), 
it fails to rule out the \emph{irrational} grid, 
although, under the strong criterion for Lorentz invariance that bans preferred reference frames,
\emph{almost all types of grid} should be ruled out.
(We write 'almost' in light of an interesting, recently identified class of 
\emph{regular} 2D lattices, called 'Lorentzian lattices' \cite{ref:SaravaniAslanbeigi2014}, which seems
to partially satisfy the 'no preferred frame' criterion:
these structures are invariant under a discrete subgroup of the Lorentz group,
while additionally offering a very good number-volume correspondence.)

The useful lesson that comes from the irrational grid counter-example is that, when
trying to characterise Lorentzianity, 
we should decouple the regularity/irregularity issue 
- with its impact on the presence/absence of preferred frames -
from other aspects, e.g. involving node degrees or path lengths (to be discussed next), 
which are nevertheless equally relevant to Lorentzianity.
Indicators for the latter aspects are unlikely to tell apart order from disorder, but this fact may indeed turn into an advantage,
when one is interested (as we are)  
in the analysis and synthesis of \emph{algorithmic} causets, which, unlike
stochastic causets, span the whole range between fully regular and (pseudo-)random patterns.
In other words, we certainly value a scenario in which some degree of Lorentzianity can be attributed 
to a causet with regular components.

%
\subsection{Longest/shortest path plots}
\label{subsect:LongestShortestPathPlots}
%

The new indicator we introduce now is still based on link counts, 
but is concerned with longest and shortest paths between nodes.
Similar to the indicators based solely on intervals, 
this one is not affected by the overall shape of the graph, or by 'border' considerations.

We shall try to characterise the extent to which a causet succeeds in reproducing a key property of Lorentzian manifolds: 
the \emph{reversed triangular inequality}.  
Consider two events $p_{1}$ and $p_{2}$ in flat Minkowski space $M^{1,3}$, with $p_{2}$ in the future lightcone of $p_{1}$.  
The (+ - - -) signature of the Lorentz metric is at the basis of the inequality, which in turn leads to the twin paradox:  
the time delay experienced by the first twin, who travels from $p_{1}$ and $p_{2}$ following a straight line (a geodesic), 
is maximum; 
any time-like trajectory $p_{1} \rightarrow x \rightarrow p_{2}$
taken by the second twin via some intermediate spacetime point $x$ registers a shorter time delay.  
In terms of Lorentz distance $L$:  
\begin{equation} 
L(p_{1}, p_{2}) \geq L(p_{1}, x) + L(x, p_{2}).
\label{eq:reversedtriangularInequality}
\end{equation}
The limit case - zero time - is registered when $p_{1} \rightarrow x$ and $x \rightarrow p_{2}$ are 
 light-like segments that form at $x$ a $\pi/2$ angle. 

When a causet $C$ is derived by sprinkling in Minkowski space $M$, 
we can approximate the Lorentz distance in $M$ between two timelike related points $p_{1}$ and $p_{2}$ 
by the length (number of edges) of the longest chain $P$ between them in $C$ \cite{ref:Myrheim1978}, 
which represents a geodesic (this correspondence is conjectured to hold also for sprinklings in curved manifolds). 
We denote this length by $lpl(p_{1},p_{2})$, for 'longest path lengh'.   
Note that all edges of path $P$ must be links, that is, edges of the transitive reduction $C_{red}$ of $C$:
if one were not a link, it could be replaced by two or more links,  yielding a longer path and conflicting with $P$ being the longest.  
Thus $C_{red}$, which is unique, codes all the necessary information for measuring the Lorentz distance between any two events.  
In the sequel we shall drop the 'red' subscript. 

How could we then statistically characterise the extent to which a causal set $C$ reproduces the inequality of equation \ref{eq:reversedtriangularInequality}?  
In light of the correspondence between $L(p_{1}, p_{2})$ and $lpl(p_{1}, p_{2})$, 
and just sticking to triangles, 
the most direct approach would be to compile statistics on, say, the difference or ratio between 
$lpl(p_{1}, p_{2})$ 
and 
$lpl(p_{1}, x) + lpl(x, p_{2})$, for all $x$ between $p_{1}$ and $p_{2}$.   
Following this line of thought, we can compute the expected gain $r$ of path 
$p_{1} \rightarrow x \rightarrow p_{2}$ over $p_{1} \rightarrow p_{2}$
when $x$ is chosen uniformly at random in a 2D or a 3D Minkowski interval 
$I[p_{1}, p_{2}]$.  
We find:
\begin{equation} %
r = \frac
{    \int_{x \in I[p_{1}, p_{2}]}      (L(p_{1}, x) + L(x, p_{2})) / L(p_{1}, p_{2}) \hspace{.1cm} dx
}
{
Vol(I[p_{1}, p_{2}])
} = 
\left\{ \begin{array}{ll}
8/9 & \mbox{(2D Minkowski interval)} \\
4/5 & \mbox{(3D Minkowski interval)}
\end{array}
\right.
\label{eq:gainRin2d3d}
\end{equation}
Note that the two constants found are indeed independent from the shape and size of the interval, 
due to Lorentz invariance.  
In the discrete setting, one can then test interval causets by finding how well they approximate these constants, 
using function \emph{lpl} and summation in place of, respectively, function $L$ and integration.  
Additionally, these constants might prove useful as dimension estimators, 
analogous to ordering fractions for the well known Myrheim-Meyer estimator.

However, we have an even simpler way to characterise the variety of path lengths between $p_{1}$ and $p_{2}$.  
Since in a Lorentzian manifold $M$ we find infinitely many trajectories between $p_{1}$ and $p_{2}$ that are shorter, 
or even \emph{much} shorter than the geodesic, we require that the finite number of alternative paths between 
two related nodes in $C$ widely range in length too, from very long to very short.   
For further simplicity, we concentrate on the extreme cases of the longest and shortest paths between event pairs.  

The interplay between these lengths is represented in a convenient compact way 
by using what we call the \emph{longest/shortest path plots}.  
The function $f_{C,s}(l)$ depicted in these plots is defined relative to a node $s$ 
- typically a root - of a causet $C$, 
refers to all and only those paths that start from $s$, 
and yields the mean shortest path length corresponding to longest path length $l$.  
Formally: 
\begin{equation} %
f_{C,s}(l) = Mean\{spl(s, x) | x \in Nodes© \wedge lpl(s, x) = l\},
\label{eq:longShortFunctionDef}
\end{equation}
where $spl(s, x)$ and $lpl(s, x)$ denote, respectively, the lengths of the shortest and longest paths from $s$ to $x$.

\begin{figure}[h]
\centering
\includegraphics[scale=.6]{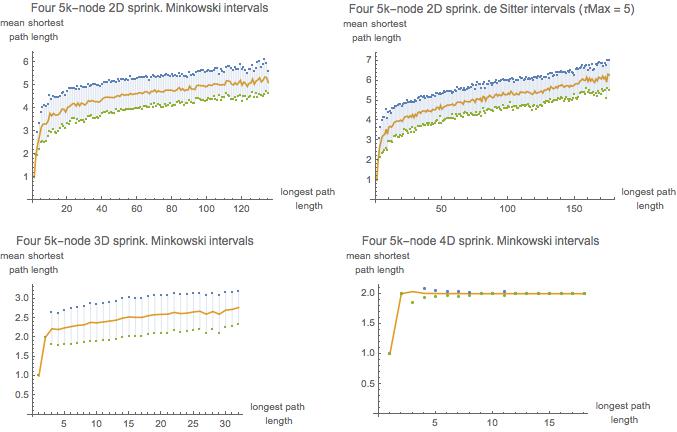}
\caption{{\footnotesize 
Longest/shortest path plots for four different types of sprinkled causet.
Each plot is obtained by considering four 5000-node interval causets of the corresponding type, 
by collecting longest and shortest path lengths from the root to all nodes,
and by computing the average shortest path length (and standard deviation) 
associated with each possible longest path length.
Upper-left: 2D sprinkled Minkowski intervals. 
Upper-right:  2D sprinkled de Sitter diamond interval, with time span (0, 5).  
Lower row: 3D and 4D sprinkled Minkowski intervals.}}
\label{fig:longShortPPfor2d3d4dSprinkIntervals}     
\end{figure}

Figure \ref{fig:longShortPPfor2d3d4dSprinkIntervals}  shows longest/shortest path plots for four different types
of transitively reduced, sprinkled causets.  
%
%
%
%
The upper-left plot refers to the 2D Minkowski case.  
For building this plot we considered four distinct 5000-node Minkowski intervals and computed, for each of them,
the longest and shortest path length pairs from the root to any node.
Then we aggregated the 20,000 pairs in the plot, according to function definition (\ref{eq:longShortFunctionDef}).

The other three plots of Figure \ref{fig:longShortPPfor2d3d4dSprinkIntervals} have been build analogously.
The upper-right plot refers to 2D sprinkled de Sitter intervals; in spite of the presence of curvature, it appears quite similar to the 2D Minkowski plot.
The lower plots refer to 3D and 4D sprinkled Minkowski intervals, and reveal two facts:
if the the number of nodes in the graph is kept constant and dimensionality increases, then
(i) the range of longest path lengths reduces, and
(ii) shortest paths associated with a fixed longest path length get shorter. 

We believe that the systematic presence of very short shortest paths between nodes that are separated by increasingly long longest paths, 
as neatly represented in these plots, reflects, in the discrete setting,  a key feature of Lorentzian manifolds.  
We shall therefore consider the presence of longest/shortest path plots \emph{qualitatively} similar to those in 
Figure \ref{fig:longShortPPfor2d3d4dSprinkIntervals} 
as a necessary requirement for causets aiming at 'Lorentzianity'. 

Let us now focus on the slow growth of the above longest/shortest path plots for the 2D cases.
\footnote{The reduced growth rate of these plots for higher dimensional cases is an interesting phenomenon that we leave for further study.}
For doing this, let us consider longest and shortest paths separately.
With slight abuse of notation, we now let $lpl(k)$ and $spl(k)$ 
define, respectively, the \emph{average} longest and \emph{average} shortest path length 
from source to sink of a sprinkled 2D Minkowski interval as a function of the number $k$ of sprinkled points
- the average being taken over large sets of intervals.

\subsubsection*{Longest path length}

In \cite{ref:BrightwellGregory91} and \cite{ref:BollobasBrightwell91} it is shown that $lpl(k)k^{-1/d}$ 
converges to some constant $m_{d}$ as the number $k$ of points sprinkled in a $d$-dimensional Minkowski interval grows to $\infty$.  
The value of $m_{d}$, however, has been calculated only for $d = 2$, and found to be $m_{2}= 2$.  
This immediately yields an asymptotic estimate of $2\sqrt(k)$ for $lpl(k)$.  

\begin{figure}[h]
\centering
\includegraphics[scale=.5]{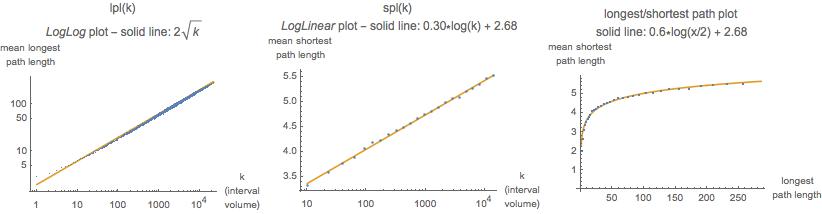}
\caption{{\footnotesize 
Left: \emph{LogLog} plot for the mean longest path length of transitively reduced sprinkled Minkowski 2D interval causets 
as a function of the sprinkling density $k$, matched against theoretical growth function $2\sqrt{k}$.  
Center: \emph{LogLinear} histogram for mean shortest path length of transitively reduced sprinkled Minkowski 2D interval causets 
as a function of the sprinkling density $k$.  Fit by natural logarithm. 
Experimental data for these two plots was derived from four 50k interval causets.  
Right: Function $f(x) = 0.3*log(x^{2}/4) + 2.68$ approximates the longest/shortest path plot for 2D Minkowski sprinkled causets 
under the assumptions $lpl(k) =  2\sqrt(k)$ and $spl(k) = 0.3log(k) + 2.68$.  
}}
\label{fig:meanShortMeanLongMeanShortLong}     
\end{figure}

The \emph{LogLog} plot in Figure \ref{fig:meanShortMeanLongMeanShortLong}-left shows the match between this theoretical prevision and experimental data.  
The latter was obtained by considering four independent 50,000-node Minkowski intervals,
then picking at random 5,000 intervals from each of them, 
then computing for each of these the pair $(k, l(k))$, 
where $k$ is the number of points in the interval (its volume)
and $l(k)$ is the length of the longest path from source to sink,
and then computing $lpl(k)$, the average of all the $l(k)$'s for a given $k$. 

 %
%
%
\subsubsection*{Shortest path length}
 
The \emph{LogLinear} plot of Figure  \ref{fig:meanShortMeanLongMeanShortLong}-center shows the mean shortest path lengths $spl(k)$ 
obtained from the same experiments used for the plot at its left.
The plot is indeed a histogram with variable bin size: the 23 data points correspond to 23 bins of geometrically growing size.  
For each bin $[k_{min}, k_{max}]$ we collect all shortest path lengths corresponding to intervals whose volume $k$ 
falls in that range, and compute their average.  
Note that this is now a \emph{LogLinear} plot: only the $x$-axis is logarithmic.  
The linear pattern indicates that $spl(k)$ has $O(log(k))$ growth.  
The matching function (solid line) is $0.3log(k)+ 2.68$.

\vspace{0.5cm}

We can now use the above results - the exact form $2\sqrt{k}$ for $lpl(k)$ and the estimate $0.3*log(k) + 2.68$ for $spl(k)$ - 
for fitting experimental longest/shortest path plots of sprinkled 2D Minkowski intervals.  
This is done via a 'hybrid' function $f$ that combines the empirical $spl(k)$ and the theoretical $lpl(k)$ by eliminating variable $k$: 
\begin{equation} 
f(x) = 0.3 log(lpl^{-1}(x)) + 2.68 = 0.3 log(\frac{x^2}{4}) + 2.68 = 0.6 log (x/2) + 2.68.
\label{eq:hybridFunctionF}
\end{equation}

The plot of function $f$ is shown in Figure \ref{fig:meanShortMeanLongMeanShortLong}-right (solid line), 
matched against the longest/shortest path plot derived from the four 50,000-node intervals
used for the first two plots.  

%
\subsubsection*{Longest/shortest path plots for the other causet classes}

\begin{figure}[h]
\centering
\includegraphics[scale=.64]{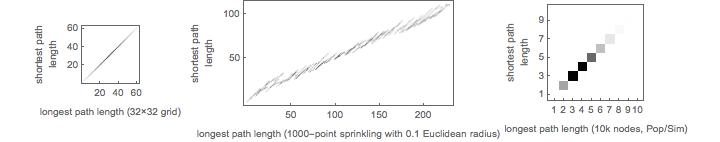}
\caption{{\footnotesize 
Longest/shortest path arrays for three causets.  
Left: 32$\times$32-node grid.  
Center: graph obtained by sprinkling 1000 points in a unit square, where edges connect points at Euclidean distance smaller than 0.1. 
Right: 10k-node pop/sim causet.
}}
\label{fig:longestShortestPathPlotFor32x32Grid}     
\end{figure}

Figure \ref{fig:longestShortestPathPlotFor32x32Grid} provides plots for  grids and pop/sim causets.  
Actually, these are longest/shortest path \emph{arrays} that present directly the raw data used for building the shortest/longest path plots as defined in equation (\ref{eq:longShortFunctionDef}): 
in each of the three  rectangular arrays the \emph{grey level} of the entry at column \emph{lpl} and row \emph{spl} 
represents the relative frequency of the $(lpl, spl)$ pairs found in the corresponding causet 
(thus, the longest/shortest path plot is obtained by simply averaging the data in the array, column by column).  

%
The longest/shortest path array for a grid graph, in which the lengths of the longest and shortest paths from the root $s$ 
to any node $x$ trivially coincide, is shown in Figure \ref{fig:longestShortestPathPlotFor32x32Grid}-left.  
The diagram provides now a particularly effective visual account of the fact that these graphs occupy 
the opposite extreme of the spectrum (maximum locality), relative to sprinkled causets (maximum non-locality). 

For the sake of comparison, we also consider a randomized version of the regular grid, 
a 'proximity' graph obtained by sprinkling points in a 2D unit square and using the Euclidean distance $d$ 
and a threshold $\delta$ for creating edges: 
a directed edge from point $p_{1}(x_{1}, y_{1})$ to point $p_{2}(x_{2}, y_{2})$  is created 
whenever $y_{2} > y_{1}$ and $d(p_{1}, p_{2}) < \delta$.  
The longest/shortest path array for such a graph, with threshold $\delta = 0.1$, is shown in 
Figure \ref{fig:longestShortestPathPlotFor32x32Grid}-center.     

Finally, Figure \ref{fig:longestShortestPathPlotFor32x32Grid}-right shows the longest/shortest path array 
for a 10k-node pop/sim causet.  
While the asymptotic behaviour of this class of causets approximates that of de Sitter sprinkled causets 
in terms of node degree distributions \cite{ref:Krioukov2012}, 
our analysis reveals, again, a remarkable difference between the two causet types, 
at the finite scale, with pop/sim causets unable to develop short paths as alternatives to long paths to the same node.  
No clue of non-locality seems to emerge in these experiments with pop/sim causets.

\begin{figure}[h]
\centering
\includegraphics[scale=.65]{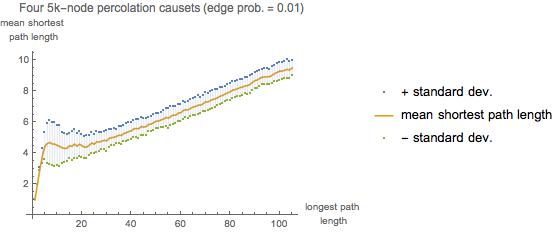}
\caption{{\footnotesize 
Longest/shortest path plot from four 5k-node percolations causets with fixed edge probability 0.01.
}}
\label{fig:longestShortestPathPlotFor5kpercolFixProb0x01}     
\end{figure}

In the case of graphs from percolation dynamics, relatively high ratios between longest and shortest path lengths can be achieved. 
This is documented by the experimental longest/shortest path plot of Figure \ref{fig:longestShortestPathPlotFor5kpercolFixProb0x01}
which, similar to the plots in Figure \ref{fig:longShortPPfor2d3d4dSprinkIntervals}, was derived from four 5k-node causets.
The edge probability we used for building them is $p = 0.01$, which yields $np = 50$ (for $n = 5000$ nodes).

%

When keeping the edge probability $p$ constant, these graphs tend to become 1D structures as $n$ grows.
This fact was already revealed by the ordering fraction spectrum of 
Figure \ref{fig:ordFractSpectraForCausetClasses}, lower-central plot,
which is characterised by the same value of $np = 50$.
As a consequence, both the longest and the shortest paths must exhibit a roughly linear growth 
with respect to the number $k$ of nodes, although the two multiplicative factors may largely differ.  
This implies that the longest/shortest path plot eventually assumes a linear character too, 
as apparent in the plot of Figure \ref{fig:longestShortestPathPlotFor5kpercolFixProb0x01}.

On the largest scale, the emergent 1D graph structure is related to the discussed phenomenon of 'posts'.
However, for $np = 50$, as in the percolation causet under discussion, the plot in Figure \ref{fig:rarityOfPosts}-left
 indicates that the expected number of posts drops below 1 as the number of nodes grows
above value $n = 150$, which means that the linear growth of the longest/shortest path plot
of Figure \ref{fig:longestShortestPathPlotFor5kpercolFixProb0x01} is \emph{not} due to the presence of posts.

In conclusion, we observe that none of the causet classes covered in 
Figures \ref{fig:longestShortestPathPlotFor32x32Grid} and \ref{fig:longestShortestPathPlotFor5kpercolFixProb0x01} 
achieves plots comparable to the one obtained for the sprinkled causets of Figure \ref{fig:longShortPPfor2d3d4dSprinkIntervals}.  
In this respect, the longest/shortest path plot indicator 
- an analytical tool of easy implementation - 
appears to exhibit a discriminative power equivalent to that of the previous node degree growth indicator, 
at least relative to the considered causet classes.

\vspace{0.5cm}

\begin{figure}[h]
\centering
\includegraphics[scale=.5]{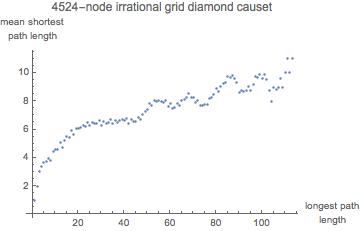}
\caption{{\footnotesize 
Longest/shortest path plots for a 4524-node diamond causet from the irrational grid.
}}
\label{fig:longShortPlotIrrationalGrid}     
\end{figure}

What about the regular, irrational grid example of the previous subsection?
Figure \ref{fig:longShortPlotIrrationalGrid} shows the longest/shortest path plot for a 4524-node irrational grid diamond causet.  

In analogy with what was observed w.r.t. node degree growth indicators,
the plot for this highly regular directed graph appears much closer to the longest/shortest path plots for 
the Lorentzian, stochastic, sprinkled causets, than to those of the other classes,+
and yet the causet violates the no-preferred-frame requirement for Lorentz invariance,  like the other grid cases.
This circumstance further confirms our previous remark on the opportunity, when dealing with Lorentzianity, 
to separate the regularity/irregularity issue from the detection of other statistical aspects 
involving node degrees or path lengths.

%
\section{New deterministic 2D causet construction techniques}
\label{sect:PermutationAnts}
%
In \cite{ref:BolognesiIJUC2010, ref:BolognesiDICE2010, ref:BologComputableUnivBook} 
the first author has described various deterministic, algorithmic causet construction techniques 
based on simple models of computation and  completely independent from an underlying manifold.  
In this section we introduce a new family of deterministic techniques that mediates between 
the manifold-dependent stochastic technique of sprinkling 
and the purely abstract (manifold-independent) algorithmic approach, in an attempt to achieve longest/shortest path plots 
-- our main indicator of Lorentzian non-locality -- 
comparable to those obtained for sprinkled causets, while retaining the benefits of the deterministic approach.  
The main benefit expected from deterministic over stochastic techniques is the emergence of structure, 
or a mix of structure and (deterministic) chaos, 
as widely discussed and shown in \cite{ref:nks, ref:BolognesiIJUC2010, ref:BolognesiDICE2010, ref:BologComputableUnivBook}.
This variety of emergent properties is also the reason why we prefer causets produced by
deterministic, dynamical, computational systems over ones, still deterministic, obtained by direct mathematical definitions,
such as the irrational grid introduced at the end of Subsection \ref{subsect:GrowthRateOfNodeDegrees}.
(Another reason for this preference is that dynamical systems appear to nicely fit a 'computational cosmology' perspective.)

We introduce a family of automata that we call \emph{permutation ants} (PA), 
in which the control unit (the 'ant') moves on a finite array $A$ of cells by short steps or jumps 
while sequentially performing simple operations on them, such as reading, writing, 
comparing, swapping cells or adding new ones.  
At each step the cell array $A$ of length $n$ contains a permutation $\pi$ of the first $n$ integers. 

The $n$-element permutation $\pi$ is directly transformed into an $n$-node causet as follows.
Each node is labeled by the pair $(i, \pi(i))$, 
which can be understood as a pair of integer coordinates; 
a directed edge $(x_{1}, y_{1}) \rightarrow (x_{2}, y_{2})$
 is created between two nodes if and only if $x_{1} < x_{2}$ and $y_{1} < y_{2}$.  
 In doing so, we are essentially still reasoning in terms of lightcones, 
 whose borders are now parallel to the cartesian axes.  
 An important difference with sprinkling is that the ant, in its journey, 
 may go back and modify previously visited sites of the growing causet. 
 
In the next two subsections we describe two types of PA automaton.  
The three pieces of information that describe them are: 
(i) the data structure on which the ant operates; 
(ii) the set $S$ of situations that are recognized by the ant; and 
(iii) the set $R$ of possible ant reactions, that depend on the situation.  
Following the approach of \cite{ref:nks}, we are interested in enumerating and exploring exhaustively 
the complete space of instances of each automaton.  
If all reactions are applicable to any situation, the size of this space is $|R|^{|S|}$.

%
\subsection{Stateful PA automaton}
\label{subsect:StatefulPAautomaton}
%
In the \emph{stateful PA automaton} the ant can be in a finite number of states, like in Turing machines. 
We shall restrict to the set of states \{0, 1\}. 

\subsubsection*{Data structure}
The ant operates on the cells of array $A$, which keeps permutation $\pi$ as described above.

\subsubsection*{Situation}
The situation is coded by 2 bits, $b_{1}$ and $b_{2}$, yielding 4 cases:

$b_{1}$ - This bit represents the current state of the ant, namely 0 or 1. 

$b_{2}$ - With the ant positioned at cell $c$ with content $x$, $b_{2}$ detects whether $x \leq c$ or $x>c$.

\subsubsection*{Reaction}
The reaction is coded by 4 bits, $b_{1}, ..., b_{4}$, yielding 16 cases.

$b_{1}$ and $b_{2}$ - these 2 bits identify 4 possible reactions, numbered from 0 to 3:

\indent \indent 0 - Swap contents of cells $c$ and $c-1$ (fails if $c-1$ does not exist);

\indent \indent 1 - Swap contents of cell $c$ and $c+1$ (fails if $c+1$ does not exist);

\indent \indent 2 - Create new cell at the right of cell $c$, with value \emph{max}+1, 
where \emph{max} is the current 

\indent \indent \indent number of cells;
	
\indent \indent 3 - Add new cell at the left of cell $c$, with value \emph{max}+1.

$b_{3}$ - This bit defines the new state of the ant.

$b_{4}$ - This bit defines the ant's move:	

\indent \indent 0 - Move one step to the left (fails if cell $c-1$ does not exist);

\indent \indent 1 - Move one step to the right (fails if cell $c+1$ does not exist).

For each of the 4 situations there is a choice among 16 reactions, 
thus there are $16^{4} = 65536$ different automaton instances, that we number by the decimal representation of the 
16 bits that characterize each of them (4 reaction bits per situation).  
We have simulated and inspected all of them, starting from initial configuration $A_{init}= (1, 2)$ 
and the ant in state $s_{init} = 0$, positioned at cell 1.  
Note that when the reaction fails - the ant attempting to access cells beyond the array limits - 
the whole computation is aborted.   
Out of the 12278 automatically selected cases that survive after 100 steps, we have manually selected two interesting cases.  
('Manual' selection consisted in displaying on the computer screen large arrays of thumbnail plots, 
each showing the ant dynamics for the first 100 or so steps, and in spotting the very few non-regular cases,
an easy and relatively fast job for the human eye.)

\begin{figure}[h]
\centering
\includegraphics[scale=.6]{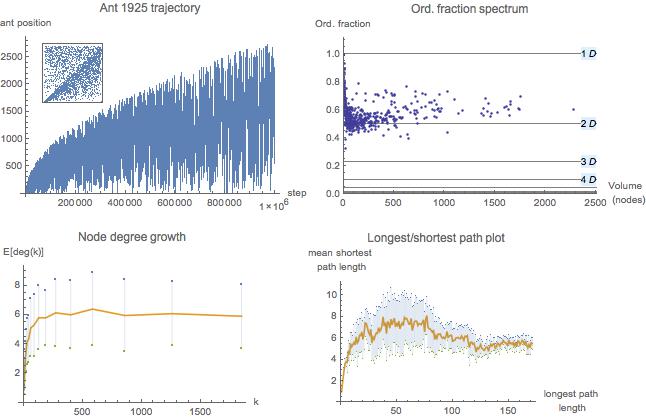}
\caption{{\footnotesize 
Stateful PA automaton n. 1925.  Ant trajectory, final permutation (inset), ordering fraction spectrum, node degree growth (counting links), longest/shortest path plot.
}}
\label{fig:statefulPAautom1925}     
\end{figure}

Automaton 1925 has the irregular and quite remarkable behaviour documented in Figure \ref{fig:statefulPAautom1925}.

The upper-left diagram plots the positions occupied by the ant on the growing array $A$ during a one-million-step computation.  
Quite remarkably, the ant keeps sweeping the full range of available cells from one extreme to the other, 
without ever entering a regular behaviour and, more surprisingly, without ever attempting to cross the boundaries.   
The final permutation is represented in the inset plot, as the set of points $(i, \pi(i))$, i = 1, ..., 2811.  
The diagonal shadow indicates that the permutation is a sort of compromise between 
the identity function and a random scattering - a mix of order and randomness.  
These two diagrams illustrate what we mean when we claim that algorithmic, 
deterministic causet construction techniques can offer surprising emergent properties 
that cannot be expected from stochastic techniques.

The upper-right plot of Figure \ref{fig:statefulPAautom1925} 
provides the ordering fraction spectrum for 500 intervals of the final causet 
- a graph with 2811 nodes and 2,456,824 edges, that decrease to 18,531 after transitive reduction - 
and suggests a Myrheim-Meyer dimension slightly less than 2D.  

The lower-left plot shows the averaged node degree (counting links) for the root nodes of intervals of variable size.
We have generated 3000 intervals $I[s, t]$ - subgraphs of the final causet -  each time picking a random source node $s$ and 
a random sink $t$ visible from $s$.  For each interval we computed the pair $(k, deg)$, where $k$ is the interval volume 
- the number of nodes - and \emph{deg} is the degree of $s$.
The histogram, with geometrically increasing bins, is derived from all these pairs: 
for each slot we plot the average \emph{deg}, 
with standard deviation, relative to all pairs for which volume $k$ falls in the slot.
The degree growth in this case is slightly weaker than that observed with the 2D Minkowski sprinkling case 
(compare with Figure \ref{fig:rootLinkGrowthInSprink2Dv8}-right).

The lower-right diagram is the longest/shortest path plot, shown with standard deviations. 
Note that we cannot exclude a priori the existence of multiple sources (and sinks) in the causet 
obtained from a permutation as described: the plot is build by considering the lengths of the longest 
and shortest link-paths from node 1 to all the nodes reachable from it, 
which may be less than the total number of nodes.
Recall that the average and the standard deviation refer to the set of shortest path lengths corresponding to
the same longest path length.

Interestingly, the longest/shortest path plot for the causet built by PA automaton 1925 
(Figure \ref{fig:statefulPAautom1925}-lower-right)
is roughly similar, numerically, to the analogous plot for 2D sprinkled Minkowski intervals 
(Figure \ref{fig:longShortPPfor2d3d4dSprinkIntervals}-upper-left).
A possibly counter-intuitive feature of this plot is that it exhibits some decreasing parts.
How is it possible that shortest path lengths decrease as longest path lengths increase?
The reason is, roughly, as follows.
When the longest paths from the root are below a certain threshold, they can only reach a portion $A$ of the causet
for which shortest paths offer limited distance reduction over the longest ones.
But as longest paths grow longer, they may eventually reach a portion $B$ of the graph
that admits fast connections from the root, via a 'short-cut region' $X$ 
that could not be exploited for reaching nodes in $A$, e.g. for the absence of paths from $X$ to $A$.
Emergent \emph{macroscopic} phenomena of this type are not possible in purely stochastic causets,
and remind us of the phenomenon of 'compartmentation' that we have observed in other algorithmic, 
pseudo-random causets \cite{ref:BolognesiDICE2010}.
Note that some decreasing segment is observed also in the longest/shortest path plot 
for the irrational grid (Figure \ref{fig:longShortPlotIrrationalGrid}).

\begin{figure}[h]
\centering
\includegraphics[scale=.6]{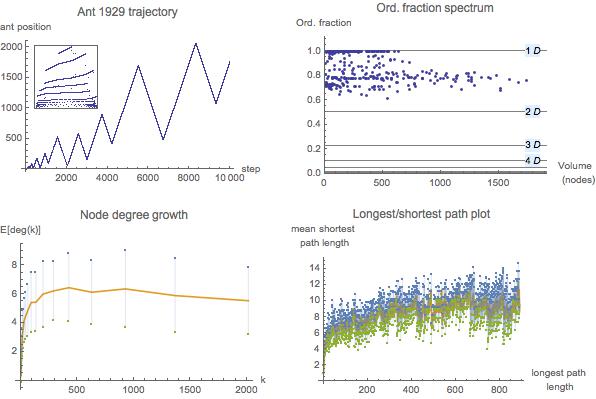}
\caption{{\footnotesize 
Stateful PA automaton n. 1929.  Ant trajectory, final permutation (inset), ordering fraction spectrum, node degree growth (counting links), longest/shortest path plot.
}}
\label{fig:statefulPAautom1929}     
\end{figure}

The stateful PA automaton 1929 has a definitely more regular behaviour, illustrated in Figure \ref{fig:statefulPAautom1929}.

%
The ant trajectory is shown in the upper-left plot for 10,000 steps (it proceeds similarly up to at least one million steps).   
The final permutation is represented in the inset plot.  
The upper-right plot shows the ordering fraction spectrum for 500 intervals of the final causet 
- a graph with 2969 nodes and 2,566,500 edges, that decrease to 5933 after transitive reduction; 
this plot differs considerably from that of the previous automaton n.\ 1925, 
providing a vague indication of a low, non-integer Myrheim-Meyer dimension.  

The lower-left diagram plots averaged node degrees (counting links) for the root nodes of intervals of variable size.
Analogous to the case of the previous automaton, the histogram uses bins of geometrically increasing size,
and reflects a slightly weaker growth than what observed in the analogous plot for 2D Minkowski sprinkling
(Figure \ref{fig:rootLinkGrowthInSprink2Dv8}-right).

In the lower-right diagram of Figure \ref{fig:statefulPAautom1929} 
we provide the longest/shortest path plot.  
Note that longest path lengths hit here record values close to 1000,
and yet the corresponding average shortest path lengths exhibit moderate growth,
yielding a plot that appears compatible with that
for 2D sprinkled Minkowski intervals (Figure \ref{fig:longShortPPfor2d3d4dSprinkIntervals}-upper-left).

%
\subsection{Stateless PA automaton}
\label{subsect:StatelessPAautomaton}
%
%
The second type of algorithm that we consider is a \emph{stateless PA automaton}.  
The control head, or ant, is now stateless, but this simplification is compensated by a slightly more complex array structure, 
and a type of reaction similar to a GOTO statement.

\subsubsection*{Data structure}
Array cell $A(i)$ is now a pair $(bit(i), \pi(i))$, where $bit(i)$ is a bit and $\pi(i)$ is the permutation element, as before.

\subsubsection*{Situation}
Coded by 2 bits, $b_{1}$ and $b_{2}$, yielding 4 cases:

$b_{1} = bit(c )$ - Cell $c$ is where the ant is currently positioned. 

$b_{2} = bit(c-1)$ - (Fails if cell c-1 does not exist).

\subsubsection*{Reaction}
Coded by 3 bits - $b_{1}$, $b_{2}$, $b_{3}$ - yielding 8 cases.

$b_{1} = 0 \Rightarrow$ Write $b_{2}$ in cell $c$ and swap cells $c$ and $c-1$ 

\indent \indent (potential failure is detected when checking the Situation, as just described).

$b_{1} = 1 \Rightarrow$ Create a new cell $(b_{2}, max + 1)$ and insert it at position c. 

\indent \indent (\emph{max} is the current number of cells).

$b_{3}$ - This bit defines the ant's move.

\indent \indent 0 - Ant does not move.

\indent \indent 1 - Ant jumps to cell $\pi(c )$ (GOTO statement; $c$ is the current ant position).
	
\vspace{0.5cm}

Reasoning as for the stateful ant, we now obtain $8^{4} = 4096$ automaton instances.  
We have simulated and inspected all of them, starting from an initial two-cell configuration 
$A_{init} = ((0, 1), (0, 2))$, with the ant positioned at cell 2.  
As with the stateful case, the computation aborts when the ant attempts to access locations beyond the array limits.   
Out of the 2876 automata that survive after 100 steps we have selected two interesting cases.  	

\begin{figure}[h]
\centering
\includegraphics[scale=.6]{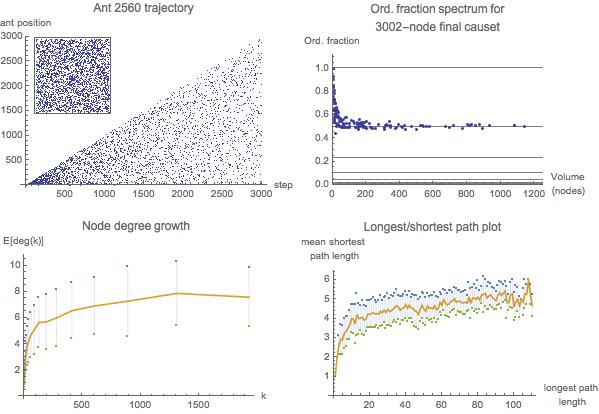}
\caption{{\footnotesize 
Stateless PA automaton n. 2560.  3000-step computation. 
Ant trajectory, final permutation (inset), ordering fraction spectrum, node degree growth (counting links), longest/shortest path plot.
}}
\label{fig:statelessPAautom2560}     
\end{figure}

Stateless PA automaton n. 2560, many copies of which are indeed found in this family, 
is remarkable for its apparently perfect random-like behaviour.  Its features are illustrated in Figure \ref{fig:statelessPAautom2560}. 
An interesting feature of these causets is that they only have one source node, since natural number 1 never moves from 
the first position in the permutation array, thus yielding causet root node (1, 1).  As a consequence,  
all nodes in the graph are visible from the root and contribute to the longest/shortest path plot.

At first sight this automaton appears as a 'perfect sprinkler'  
since its ordering fraction spectrum, 
node degree growth
and longest/shortest path plot 
are basically indistinguishable from those of sprinkled 2D Minkowski causets;
a closer analysis, however, reveals some slight departure from pure sprinkling.
We have estimated function $lpl(k)$ three times, relative to three distinct final causets obtained by letting the
automaton run, respectively, for $n = 1000, 2000, 3000$ steps.  
For each of these causets we have collected hundreds of (\emph{vol}, \emph{lpl}) pairs, 
where \emph{vol} is the volume of a subinterval $I[s, t]$ of the causet - $s$ being the causet root -
and \emph{lpl} is the length of the longest
path from $s$ to $t$, and then we have computed the best fit of these data
against function $a*k^{b}$, keeping in mind the reference values $a = 2$ and $b = 1/2$ for 2D sprinkled Minkowski intervals 
(see also Figure \ref{fig:meanShortMeanLongMeanShortLong}-left).

We found that, as the number $n$ of nodes of the causet grows ($n = 1000 \rightarrow 2000 \rightarrow 3000$),
fitting parameters $a$ and $b$ tend to drift away from the above reference values: $a = 1.95 \rightarrow 1.75 \rightarrow 1.50$,
$b = 0.50 \rightarrow 0.52 \rightarrow 0.54$.  
It would be interesting to find whether this tendency persist with much higher values of $n$ - an analysis which is unfortunately limited by
the severe computational bottleneck of transitive reduction.
%


Another interesting feature of this automaton is that its definition can be greatly simplified without affecting its behaviour.
We noticed that the final causet built in $n$ steps has $n+2$ nodes.  
This means that, starting from the two-cell initial array, the ant reaction is always of one type: create a new cell at each step.  
A closer look at the behaviour of the computation reveals that several options in the algorithm 
are never used by specific instance n.\ 2560.  
In particular, even the bits that decorate array cells can be eliminated!  
We can then provide a very concise algorithm that performs exactly the same computation.  
Here is the tiny \emph{Mathematica} code for the simplified ant step: 

{\footnotesize 
\begin{verbatim}
      step[{array_, pos_}] := {Insert[array, Length[array] + 1, pos], array[[pos]]}
\end{verbatim}
}

The above function 'step' takes a pair $\{array, pos\}$, where 
\emph{array} represents a permutation of the first $n$ positive integers
and \emph{pos} is an integer identifying a position in it, 
and returns a new pair $\{array', pos'\}$, where 
\emph{array'} is the result of inserting value $n+1$ at position \emph{pos} of \emph{array}
(so that \emph{array'} is now a permutation of the first $n+1$ positive integers),
and \emph{pos'} is the integer found at position \emph{pos} of \emph{array}.
By iterating the function call 3000 times with initial array (1, 2) and initial ant position 2, 
one obtains exactly the same results of Figure \ref{fig:statelessPAautom2560}.  

We believe that this concise randomization algorithm might be of interest, independent of the application to causets;
it would be interesting to further investigate
its statistical qualities
or its possible relation with other such generators.

\begin{figure}[h]
\centering
\includegraphics[scale=.58]{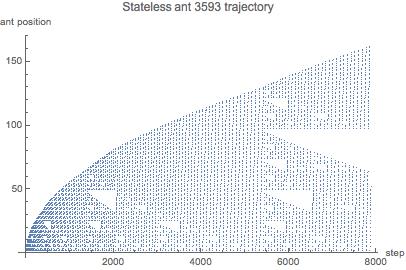}
\caption{{\footnotesize 
Ant trajectory for stateless PA automaton n. 3593.
}}
\label{fig:statelessPAautom3593}     
\end{figure}

In Figure \ref{fig:statelessPAautom3593} we present just the ant trajectory for instance 3593 
of the stateless PA automaton. 
%
%
The causet obtained by this permutation is one-dimensional, thus 
 uninteresting in terms of 'Lorentzianity' and non-locality.  
The reason for presenting the plot is to show yet another instance of the typical, 
triangle-based self-similar patterns that emerge in cellular automata and other models, as widely illustrated in \cite{ref:nks}: 
it is precisely this richness of emergent phenomena, from regular to pseudo-random patterns, 
that motivates our interest for algorithmic causets.

%
\section{Conclusions}
\label{sect:Conclusions}



In this paper we have introduced statistical indicators for the assessment and comparison of causal set classes, 
meant to focus specifically on their 'Lorentzianity', intended as the manifestation, 
in the discrete setting, of the 'non-local' nature of Minkowski space, as implied by the Lorentz distance.  
Our experimental work, based on extensive computer simulations, has led us to a number of conclusions.

First, we have established a clear distinction between indicators based on transitively closed (Section \ref{sect:CountingEdges}) 
and transitively reduced (Section \ref{sect:CountingLinks}) causets.  
In spite of their usefulness for Myrheim-Meyer dimension estimation, 
indicators of the first type (at least those we have considered in this paper) fail to discriminate between cases 
as different as the highly non-local sprinkled causets and highly local directed graphs such as a regular grid,
thus proving inadequate for characterising 'Lorentzianity'.
Furthermore we have shown that a power-law distribution of node degrees is not a rare property of transitively closed causets, 
and does not single out sprinkled de Sitter causets \cite{ref:Krioukov2012} as a special case of discrete spacetime, 
being present, for example, in percolation causets and in what we called sprinkled Minkowski cylinder 
(Figure \ref{fig:TrCloNodeDegsRotationalCausets}).  
Note that, whenever the estimation of causet node degrees is obtained by measuring lightcone areas 
in the embedding manifold, as done in \cite{ref:Krioukov2012}, 
the analysis is inevitably bound to address only the transitively closed version of the causet, 
with the limitations just mentioned.  

Considering transitively reduced graphs proves more useful, although, unfortunately, 
transitive reduction is computationally costly \cite{ref:AhoGarUll72}, 
being $O(|V| |E|)$ for a directed acyclic graph $G(V, E)$ 
- this has been the main computational bottleneck of our investigation.  
We have proved, analytically and experimentally, that the node degrees of a transitively reduced sprinkled Minkowski interval causet 
exhibit $O(log(k))$ growth in 2D and $O(k^{1/3})$ growth in 3D, 
where $k$ is the number of sprinkled nodes, 
confirming what concisely stated, in slightly different terms and without proof, in \cite{ref:bombelliReply88}.

Still in the context of transitively reduced causets, we have introduced a new indicator: 
the longest/shortest path plot.  
This simple but effective visual indicator is designed to directly reflect the interplay 
between longest and shortest link-paths, whose length differences grow very large 
in causets derived from sprinkling in Lorentzian manifolds.  
We have shown that the very slow growth that these plots exhibit for 
sprinkled Minkowski and de Sitter intervals
is not observed in the other considered causet classes, for which the growth is linear 
(Figures \ref{fig:longestShortestPathPlotFor32x32Grid}  and \ref{fig:longestShortestPathPlotFor5kpercolFixProb0x01}).

We found that the node degree growth and the longest/shortest path plot have analogous discriminative power,
at least for the considered causet classes, although the latter was meant to reflect more directly a peculiar feature
of the Lorentzian metrics, one related to the reversed triangular inequality.
Indeed, an attractive item for further research would be to try and decouple these two indicators, 
checking whether specific causet classes exist for which they provide divergent responses. 

More work is also necessary to shed further light on the relation between longest/shortest path plots and causet
dimensionality, in light of the experimental observation that these plots tend to flatten as the latter increases.

We have then introduced two new classes of deterministic, algorithmic causets, 
built by stateless and stateful PA (Permutation Ant) automata, and have identified a 'perfect sprinkler' 
- a deterministic 'ant' able to build
a causet that appears, under the lens of the longest/shortest path length indicator, 
indistinguishable from a sprinkled causet.  
For this 'ant' we have also provided an extremely concise implementation - one line of code.  
A peculiarity of this approach is that, unlike in 'cumulative', stochastic sequential approaches, 
the topology of the growing causet can be modified by the ant at any location, at any step.

The ability to mimic the randomness of sprinkled causets by a deterministic approach,
such as our permutation-based 'ant', is interesting, and tells something about the power of the automaton;
but it is not too surprising, 
in light of the rather direct correspondence between permutations and sprinkling,
and of the widely known fact that many simple models of computation can produce 'deterministic chaos'
(in \cite{ref:nks} this ability is conjectured to be a clue for computational universality).
Quite oppositely, the main expected advantage of algorithmic causets over stochastic ones
is the emergence in the former of some regularity, or a mix or regularity 
and pseudo-randomness.
The few examples we have provided here indicate that 
progress in this direction is still at an early stage,
and that
more work is needed for finding algorithmic causet classes offering the mentioned mix
while achieving a good performance in terms of longest/shortest path plots, i.e. Lorentzian non-locality.
Some promising examples, obtained very recently by a 4-state automaton similar to 
the one presented here, are described in \cite{ref:BolognesiAdamatzkyChapter2016}.

When causets exhibit some regularity, 
preferred frames of reference appear that violate Lorentz invariance, 
as with the discussed regular grids
(the 'Lorentzian lattices' \cite{ref:SaravaniAslanbeigi2014}
mentioned at the end of Subsection \ref{subsubsect:irrationalGrid} are an interesting exception). 
In this respect, node degree growth rate and longest/shortest path plots (Section \ref{sect:CountingLinks})
offer the advantage of providing useful information on a weaker form of Lorentzianity,
independent from concerns on preferred frames.

The specific causets produced by deterministic Permutation Ants are always embeddable in 2D Minkowski space,
like those produced by stochastic sprinkling.
Of course this is not a general property of deterministic causets.  
But there is no point in insisting on direct embeddability:
quantum oscillations at ultra-low spacetime scales may conflict with this requirement,
and such causets may still turn out to be embeddable in a coarse-grained form.

There exists a growing body of papers on causet embeddability, manifoldlikeness and 'Lorentzianity'.  
The objective of these efforts is to provide  tools for the reconstruction, whenever possible, 
of continuum information from the discrete structure of the causet; 
investigated properties include dimensionality \cite{ref:MeyerThesis1989}, 
time-like and space-like distance \cite{ref:Myrheim1978, ref:RideoutWallden2009}, 
curvature \cite{ref:BenincasaDawker2010}.  
(Unfortunately, the relative simplicity of the definitions and tools for these properties in the continuum 
is easily lost when transposing them in the discrete setting.)

The specific problem of characterising locality 
- local regions that are small with respect to the curvature scale -  
is addressed in \cite{ref:GlaserSurya2013}, where the reader can also find additional references for the other mentioned properties.  
The problem here is that in a causet obtained by sprinkling in a Lorentzian manifold, 
two intervals $I[p, q]$ and $I[p, r]$ with the same root and the same volume 
may largely differ in the depth of the neighborhood that they span, 
to the point that one may fall below and the other above the scale of curvature, thus qualifying, respectively, as local and non-local.  
As observed in \cite{ref:GlaserSurya2013}, 
"\emph{from the continuum perspective small, or local neighbourhoods are essential to several geometric constructions and are key to the conception of a manifold.}"  
The technique elaborated in \cite{ref:GlaserSurya2013} is based on detecting the distribution of 
$m$-element intervals in flat spacetime, as a function of $m$, 
and in using these profiles as a benchmark for assessing the flatness/locality of a causet region.   
Subtle correlations might exist between these refined locality/flatness detectors and our longest/shortest path plots.  
However, our plots are meant to detect an important clue for Lorentzianity 
- the potential presence of a wide gap between longest and shortest paths between the same two points -  
regardless of whether the inspected region is below or above the curvature scale (if any).  
Furthermore, a longest/shortest path plot that grows very slowly can still be taken, in our opinion, 
as an indicator of non-locality (with possible overloading of the term), 
even when dealing with a flat and 'small' region of the graph.

It would also be interesting, moving above the ground level of spacetime, to look for other cases in Nature 
(e.g. in the biosphere) where 'non-locality', as revealed by our plots, plays some role, and to study how it is implemented.  
For example, the evolution of plant leaf venation has led to patterns that optimize hydraulics and tolerance to cuts, 
and involve long paths for nutrient transportation between points at a short distance from each other.

\subsubsection{Acknowledgement}
The first author expresses his gratitude to Marco Tarini for useful discussions and experiments on shortest and longest paths in sprinkled causets, and to an anonymous referee 
for proposing the 'irrational grid' example,
for suggesting improvements in the calculations of expected node degrees in Minkowski 2D and 3D sprinkled intervals,  
and for suggesting additional references on causet manifoldlikeness.  
This work was supported by CNR, Consiglio Nazionale delle Ricerche/Istituto ISTI/FMT Laboratory.




\newpage


\section*{Appendix A - Node degree distribution for (transitively closed) percolation causets}

\begin{figure}[h]
\centering
\includegraphics[scale=.6]{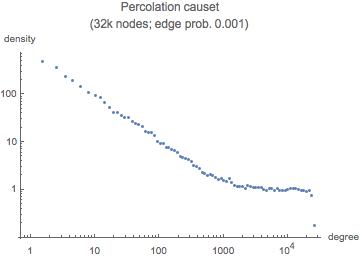}
\caption{{\footnotesize Node degree density for transitively closed percolation causet.  Slots of geometrically increasing width have been used, and
the node counts for each slot have been normalised accordingly.}
}
\label{fig:TrCloNodeDegsPercolation}     
\end{figure}

The node degree density of Figure \ref{fig:TrCloNodeDegsPercolation} refers to the transitive closure of 
a 32k-node percolation causet in which 
the probability of finding an edge between any two nodes $i$ and $j$ is $p(i, j) = 0.001$.    
(Recall that we always deal with \emph{out}-degrees.)
A power-law seems to be still in action here, up to about degree 1000,
while a flat plot segment reveals a constant density for higher degrees.

Note that we are considering a \emph{full} causet, not an interval, hence we cannot compare this plot with those
of Figure \ref{fig:TrCloNodeDegsIntervalCausets}.
Comparison with the plots of Figures \ref{fig:TrCloNodeDegsRotationalCausets} and \ref{fig:TrCloNodeDegsPopSim},
covering \emph{full} causets with rotational symmetry is perhaps more justified, although percolation causets do not share 
that symmetry.

The reason for the two-fold character of the density plot 
- power law for low degrees, flat region for higher degrees - 
is better understood by considering the influence 
on the graph structure
of parameter $np$, where $n$ is the number of nodes and $p$ is the edge probability.
The importance of this parameter is well established for the Gilbert model of random graphs
to which raw percolation graphs belong.  
In short (see  \cite{ref:Bollobas84} for more details) 
the value $np = 1$ marks a change of phase in the 
sizes of the connected components of the graph. 
When $np < 1$, the graph is fragmented into many small connected components, 
the largest of which has only O(log $n$) nodes.
When $np =1$ the size of the largest component is, with high probability, proportional to $n^\frac{2}{3}$.
Only when $np >1$ does a single \emph{giant component} emerge, with size proportional to $n$,
while the other components are still of O(log $n$) size.

\begin{figure}[h]
\centering
\includegraphics[scale=.47]{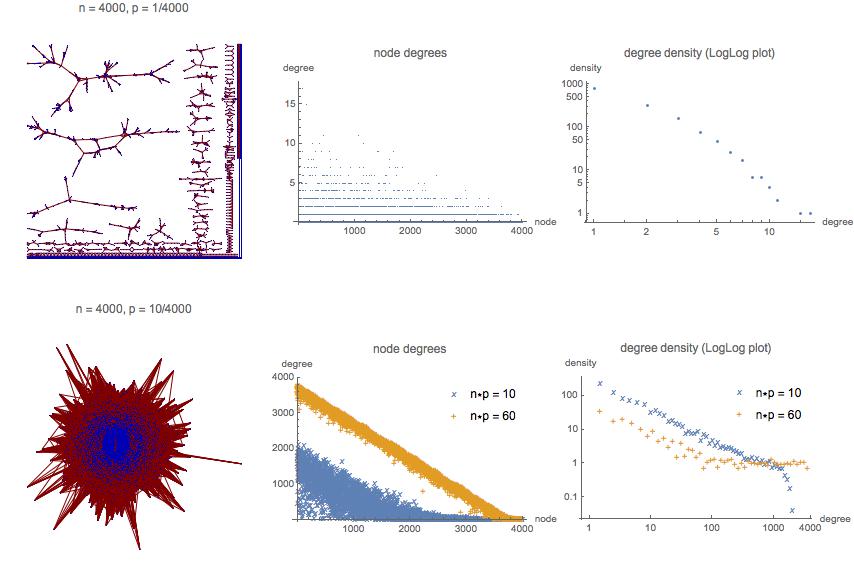}
\caption{{\footnotesize Node degree densities for transitively closed percolation graphs with three different values of parameter $n p$,
where $n = 4000$ is the number of nodes and $p$ is the edge probability. 
Upper row:  $n p$ = 1.  In this critical case the raw graph and its transitive closure (only the latter is shown)
are disconnected, thus node degrees are dramatically reduced ($\leq 17$).   The apparent lines
at two borders of the upper-left plot are indeed formed by densely packed one-node components. 
The central plot shows the degrees of individual nodes, while the plot at the right provides degree density, roughly following a power law.
Lower row: $n p$ = 10 and 60.  
The raw graph and its transitive closure are now connected 
(only the transitively closed graph for $np = 10$ is shown).
Individual node degrees and degree densities are also illustrated, for both values of $np$, in the central and r.h.s. plots.}
}
\label{fig:cloPercolDegsAdditionalPlots}     
\end{figure}

The fragmentation of the graph is illustrated in the upper left diagram of Figure \ref{fig:cloPercolDegsAdditionalPlots},
that represents the transitive closure of the 4000-node graph ($n$ = 4000) 
obtained for the critical value $np = 1$ ($p = 1/4000$).

Given this fragmentation, node degrees are limited:
the remaining two plots of the upper row of Figure \ref{fig:cloPercolDegsAdditionalPlots}
- degrees of individual nodes, ordered by increasing label, and degree density - reveal that
the maximum degree achieved is 17, and that the degree density roughly follows a power law.

As $np$ grows beyond unit value, up to the values 10 and 60 considered
in the lower row of Figure \ref{fig:cloPercolDegsAdditionalPlots},
the graph rapidly becomes connected 
(note that only the graph for case $np = 10$ is shown). 
This fact dramatically boosts node degrees of the transitive closure.
The central diagram combines the node degree plots for the two cases: the nodes with very low index
have degrees that roughly reach, respectively, values 2000 and 4000.  

The presence of a flat region with constant value in the degree density plots 
(Figure \ref{fig:TrCloNodeDegsPercolation} and Figure \ref{fig:cloPercolDegsAdditionalPlots}-lower-right)
directly reflects the presence of the linearly decreasing strip of points in the plots for individual node degrees,
particularly apparent for the $np = 60$ case (Figure \ref{fig:cloPercolDegsAdditionalPlots}-lower-middle).
The reader can also easily work out the reason why that constant value is around 1.

Indeed, the expected degree of the node with index $x$ decreases linearly with $x$ also in the raw graph $R$, i.e. before
transitive closure.
This is because the degree of $x$ is the sum of the 0-1 outcomes of $(n-x)$
independent experiments, each with probability $p$ of \emph{success} (outcome = 1), yielding
an expected degree $\mu = (n-x)p$.

The power-law component in the degree density plots for high values of $np = 10$ and $np = 60$ 
(Figure \ref{fig:cloPercolDegsAdditionalPlots})
is due to the fact that 
the graph fragmentation phenomenon and its effect on node degrees is still lurking, 
even when $np \gg 1$ and the graph as a whole is connected.  
For the given edge probability $p$, there is always a small, $m$-node subgraph $M$ consisting of the last, highest indexed $m$ nodes, for which the critical parameter $mp$ is of the order of 1.
The contribution of $M$ to the out-degree distribution is that of a disconnected graph 
- yielding a power-law as in the upper-right plot of Figure \ref{fig:cloPercolDegsAdditionalPlots} -
although the subgraph is still connected to the rest of the graph via edges that emanate from nodes with smaller indices.


\end{document}